\documentclass[twocolumn,showpacs,preprintnumbers,nofootinbib,prd,superscriptaddress,10pt]{revtex4-1}

\usepackage{bm,graphics,graphicx,epsfig,soul,latexsym,hyperref,mathrsfs,multirow,natbib,url}
\usepackage[dvipsnames]{xcolor}
\usepackage{amsmath}
\usepackage{amsmath,amssymb}
\usepackage{setspace}

\newcommand{\bmSeffp}{{\bm{S}_{0}^+}}
\newcommand{\bmSeffm}{{\bm{S}_{0}^-}}

\usepackage[english]{babel}
\usepackage[normalem]{ulem}
\definecolor{darkgreen}{rgb}{0,0.5,0}

\hypersetup{
    unicode=false,          
    pdftoolbar=true,        
    pdfmenubar=true,        
    pdffitwindow=false,     
    pdfstartview={FitH},    
    pdftitle={My title},    
    pdfauthor={Author},     
    pdfsubject={Subject},   
    pdfcreator={Creator},   
    pdfproducer={Producer}, 
    pdfkeywords={keyword1} {key2} {key3}, 
    pdfnewwindow=true,      
    colorlinks=true,       
    linkcolor=red,          
    citecolor=cyan,        
    filecolor=magenta,      
    urlcolor=darkgreen,           
    linktocpage=true
}

\allowdisplaybreaks
\begin{document}

\title{Spin effects in spherical harmonic modes of gravitational waves from eccentric compact binary inspirals}
\author{Kaushik Paul}
\email{kpaul.gw@gmail.com}
\affiliation{Department of Physics, Indian Institute of Technology Madras, Chennai 600036, India}
\affiliation{Centre for Strings, Gravitation and Cosmology, Department of Physics, Indian Institute of Technology Madras, Chennai 600036, India}

\author{Chandra Kant Mishra}
\email{ckm@iitm.ac.in}
\affiliation{Department of Physics, Indian Institute of Technology Madras, Chennai 600036, India}
\affiliation{Centre for Strings, Gravitation and Cosmology, Department of Physics, Indian Institute of Technology Madras, Chennai 600036, India}
\begin{abstract}
We compute the leading and subleading spin effects through the second post-Newtonian order (2PN) in spherical harmonic modes of gravitational waveforms from inspiralling compact binaries in \textit{noncircular} orbits with \textit{nonprecessing} components. The two spin couplings, linear-in-spin (spin-orbit) and quadratic-in-spin (spin-spin), that appear in 2PN waveforms are computed with desired accuracy and explicit expressions for relevant modes are derived. The modes that have spin corrections through 2PN include $(\ell, |m|)$=$((2,2),\,(2,1),\,(3,3),\,(3,2),\,(3,1),\,(4,3),\,(4,1))$ modes. Additionally, two $m=0$ modes---$(2,0)$ and $(3,0)$---also contribute to the 2PN order. Closed-form expressions for these modes for compact binaries in general orbits as well as in elliptical orbits are provided. While the general orbit results can be used to study signals from binaries in orbits of arbitrary shape and nature, elliptical orbit results are applicable to systems with arbitrary eccentricities. We also express the elliptical orbit results as leading eccentric corrections to the circular results. Our prescription represents, the first, fully analytical treatment that combines spins, eccentricity and higher modes together and completes computation of spin effects through 2PN order. These should find immediate applications in inspiral-merger-ringdown modeling for eccentric mergers including the effect of nonprecessing spins and higher modes as well as in parameter estimation analyses employing an inspiral waveform.
\end{abstract}
\date{\today}
\maketitle
\section{Introduction} 
\label{sec:intro}
The ground-based network of gravitational-wave (GW) detectors involving the LIGO \cite{LIGOScientific:2014pky} and Virgo \cite{VIRGO:2014yos} detectors has detected so far nearly 90 compact binary mergers \citep{LIGOScientific:2018mvr, LIGOScientific:2020ibl, LIGOScientific:2021usb, LIGOScientific:2021djp}. While the majority of these events are inferred to be binary black hole (BBH) mergers, two binary neutron star~\citep{LIGOScientific:2017vwq,LIGOScientific:2020aai} and two neutron star-black hole ~\citep{LIGOScientific:2021qlt} mergers have also been observed. While compact binaries involving neutron stars and stellar mass black holes (BHs) will continue to dominate the frequency bands observable from ground $[ $above 1-10Hz; observable by planned (LIGO A+~\cite{Shoemaker:2019bqt, McClelland:T1500290-v3}, Voyager~\cite{McClelland:T1500290-v3}) and proposed future $[$Cosmic Explorer (CE)~\cite{Dwyer:2014fpa}, Einstein Telescope (ET)~\cite{Punturo:2010zz} detectors configurations, intermediate mass BHs and supermassive BHs will dominate the observation bands accessible only from space $[$in deci-Hz band (DECIGO~\cite{Sato_2017}) and in milli-Hz band (LISA~\cite{amaro2017laser})$]$. Detection of GWs from these sources involves cross-correlating the detector data with a collection of simulated copies (called templates) of expected signals. This template-based search method is referred to as matched filtering \cite{Cutler:1994ys, Poisson:1995ef, Krolak:1995md} and its reliability critically depends on the closeness of the templates to the true signal buried in the detector's noisy data. The requirements on the accuracy of these templates is quite stringent for detection (mismatches with the true signal should be no larger than $\sim3$\%) and for subsequent parameter estimation analyses (no larger than $\sim$1\%). (See for instance Ref.~\cite{LIGOScientific:2016vbw} for details on search methods currently in use for analyzing data from LIGO and Virgo detectors.)
\vskip 5pt
Templates that are currently used in performing GW searches assume binaries in inspiralling circular orbits owing to the expected circularisation of the binary \cite{PhysRev.136.B1224, PhysRev.131.435}. Although, partly it is also due to the absence of appropriate signal models (and of pertinent search strategies\footnote{That optimally manage the cost of adding another direction (orbital eccentricity) to the search parameter space.}) including the effects of orbital eccentricity. Moreover, compact binaries formed via dynamical interactions in dense stellar environments or (if the binary is part of a stable hierarchical triple system) through Kozai-Lidov processes \cite{Kozai:1962zz,LIDOV1962719} are expected to be observed in ground-based detectors with residual eccentricities ($e_0$)~$\gtrsim0.1$ \cite{Salemi:2019owp}. In fact, some of the recent analyses \cite{GW190521-properties, Kimball:2020qyd, Romero-Shaw:2021ual, Romero-Shaw:2022xko, OShea:2021ugg} show support for the presence of eccentricity in the observed data for BBH events. While it may be still possible to detect systems with residual eccentricities, $e_0\lesssim0.1$ (at 10Hz), in ground-based detectors by employing quasicircular templates, systems with larger eccentricities will require constructing templates including the effect of eccentricity\,\cite{Brown:2009ng, Huerta:2013qb}. Additionally, as argued in Ref.~\cite{Favata:2021vhw}, the presence of even smaller eccentricities ($e_0 \sim$ 0.01-0.05) may induce systematic biases in parameter estimation analyses focusing on extracting source parameters. Moreover, the next generation ground based detectors, CE~\cite{Dwyer:2014fpa} and ET~\cite{Punturo:2010zz}, with improved low frequency sensitivity compared to the current generation detectors, should make confident as well as frequent observations of eccentric systems \cite{Lower:2018seu, Tibrewal-etal-2021}. 
\vskip 5pt
It may be worth noting that regardless of the availability of eccentric waveforms for compact binary searches, current search strategies will need thorough revisions in order to perform template based eccentric searches. In fact, current template-based searches employing circular templates too ignore a number of important physical effects such as those due to precessing spins and nonquadrupole modes to manage the cost of the analysis \cite{LIGOScientific:2016vbw}. More complete models are used in subsequent analyses to assess the impact of different physical effects on measurements of important source parameters such as masses and spins and/or to measure of the said effect in a parameter estimation analyses. And hence, even though eccentric models being developed today may not be used in conducting eccentric searches immediately, these may be used, in injection analyses, to check the suitability of circular templates in analyzing the data, or in a parameter estimation analysis employing eccentric waveforms, to measure eccentricity. The latter, in the context of ground based detectors, has important astrophysical implications, as eccentricity can be used as a discriminator for  the binary's formation channels (see \cite{Salemi:2019owp} for a discussion). These considerations have greatly motivated the (ongoing) efforts worldwide in the direction of waveform development including the effect of eccentricity.
\vskip 5 pt
In general, a compact binary merger is characterized by a set of intrinsic (component masses and spins) and extrinsic parameters (binary's location and orientation). Additionally, if the binary is in a noncircular orbit or has at least one neutron star, eccentricity and tidal deformability parameters may also become important. While the tidal effects only become important during late stages of binary evolution for systems with at least one neutron star, eccentricity may contribute through the entire inspiral stage depending upon its value when the signal enters the detector's sensitive band. Further, the presence of eccentricity induces modulations in signal waveforms typical of eccentric systems (see for instance, \cite{Chattaraj:2022tay, Hinder:2017sxy}).  The larger the eccentricity is, the larger these modulations become and hence their impact on detection and parameter estimation. Additionally, contributions from nonquadrupole modes become relevant for systems with nonequal mass components and/or if the binary's orbit is not face-on.\footnote{Face-on systems are those which have orbital plane perpendicular to our line of sight.}Each of these modes are further subject to eccentricity induced modulations indicative of the fact that two effects (of orbital eccentricity and nonquadrupole modes) are inseparable when modeling eccentric systems \cite{Chattaraj:2022tay} and neglecting one or both may lead to significant biases or worse nondetection \cite{Rebei:2018lzh}.
\vskip 5pt
There have been numerous efforts in the past focusing on including eccentricity in inspiral waveforms from compact binary mergers~\citep{Mishra:2015bqa, Moore:2016qxz, Tanay:2016zog, Boetzel:2019nfw, Ebersold:2019kdc, Konigsdorffer:2006zt, Moore:2019xkm}. While these efforts model eccentricity together with the effect of nonquadrupole modes, these assume binary components are nonspinning. 
Even though the effect of spins (including the effect of spin precession) has been modeled within the post-Newtonian formalism~\citep{ Arun:2008kb,Hartung:2011te, Marsat:2012fn, Levi:2015uxa,Bohe:2012mr,Bohe:2013cla,Marsat:2013caa,Bohe:2015ana,Blanchet:2011zv, Klein:2013qda, Chatziioannou:2013dza, Henry:2022dzx}, a combined treatment including spins and eccentricity is largely absent. Although, there have been efforts such as those of \citep{Kidder:1995zr, Buonanno:2012rv, Majar:2008zz, Klein:2010ti, Klein:2018ybm, Vasuth:2007gx, Klein:2021jtd} that attempt to address this concern to some extent. For instance, spin contributions to the 2PN order in a radiation field ($h_{ij}$) for binaries in general orbits are available and can be connected to GW polarization and subsequently to the spherical harmonic modes of gravitational waveforms using standard formulas (see for instance Ref.~\cite{Blanchet:2008je}). Spherical harmonic modes can also be directly computed using the inputs from the PN theory following the approach of \cite{Kidder:2007rt}. Typically it is this form of gravitational waveforms that is extracted in numerical relativity (NR) simulations. Additionally, they can readily be combined using the usual basis of spherical harmonics to obtain GW polarizations which are typically useful for data-analysis purposes. In this work, we compute spin contributions through 2PN order in spherical harmonic modes of gravitational waveforms following the approach of \cite{Kidder:2007rt} using the inputs developed within the framework of the post-Newtonian, multipolar-post-Minowskian (PN-MPM) approach \cite{Blanchet:2013haa, Buonanno:2012rv}. 
\subsection{Summary of the current work}
\label{sec:summary}
The current work can be seen as an extension to earlier efforts of Refs.~\cite{Arun:2008kb, Buonanno:2012rv, Mishra:2015bqa}. On one hand Refs.~\cite{Kidder:1995zr, Buonanno:2012rv, Arun:2008kb} together present the computation of spin effects in GW modes contributing up to 2PN for binaries in inspiralling circular orbits. On the other hand, Ref.~\cite{Mishra:2015bqa} lists explicitly nonspinning contributions to each mode for binaries in noncircular orbits. Typically, in PN literature, generalizations to circular orbit results are outlined under two separate heads. First, one computes results for general orbits characterized by a set of position and velocity variables. In the next step, one then specializes to a specific orbit (an ellipse, a hyperbola or a parabola) with the aid of relations that connect these general orbit variables to the parameters of the special orbit one intends to study. This presentation style is followed here too. 
\vskip 5pt
Note that, through the 2PN order, only two kinds of spin terms appear in mode expressions; \textit{linear-in-spin} terms that arise due to couplings between individual component spins and the orbital angular momentum and referred to as spin-orbit (SO) terms, and the second, \textit{quadratic-in-spin} terms due to interactions between individual spin vectors referred to as spin-spin (SS) terms \cite{Kidder:1995zr, Arun:2008kb, Buonanno:2012rv}. Additionally, not all modes contributing to the gravitational waveforms to 2PN have spin contributions. Modes that have spin contributions are ($\ell$, $|m|$)=($(2,2)$,\,$(2,1)$,\,$(3,3)$,\,$(3,2)$,\,$(3,1)$,\,$(4,3)$,\,$(4,1)$) and we choose to present only these here. In addition, $(2,0)$ and $(3,0)$ modes are also included as these are the two $m=0$ modes that contribute to 2PN order. Nonspinning contributions to modes which do not have spin dependencies through 2PN can be found in \cite{Mishra:2015bqa} both for general and elliptical cases. Further, in the current work we restrict ourselves to nonprecessing (no spin precession) binaries and choose to present our general orbit results in terms of two different combinations of spin parameters following the approach of Refs.~\cite{Buonanno:2012rv} and~\cite{Mishra:2016whh}.
\vskip 5pt
We first list our general orbit results in terms of a set of radial ($r, \dot{r}$) and angular ($\phi, \dot{\phi}$) position and velocity variables in Sec.~\ref{sec:genOrb_hlm}. These results are applicable to compact binaries in any noncircular orbits although these results are valid only for binaries with \textit{nonprecessing} spins. Next, we specialize to the case of elliptical orbits using the quasi-Keplerian representation of \cite{Klein:2010ti, Klein:2018ybm} for spinning compact binaries. While the representation can be used to describe binaries with even \textit{precessing} spins, we restrict ourselves to the case of non-precessing systems. We list our results for the elliptical case in Sec.~\ref{sec:qk} as well as in Appendix \ref{sec:qk_arb} and also as Supplemental Material~\cite{kp:suppl}.
\vskip 5pt
We would like to highlight here that the results presented in Appendix~\ref{sec:qk_arb} are applicable to elliptical orbits of arbitrary eccentricity and hence are probably the most useful results of the current paper. However, these have been pushed to the appendix to avoid any disruption of the flow of the paper as these expressions run over a couple of pages. Instead in Sec.~\ref{sec:qk} we list these results as leading eccentric corrections to the circular case by expanding our arbitrary eccentricity results treating eccentricity as a small parameter. While these should only be applied to orbits with small eccentricity, these significantly improve clarity of the presentation as the circular limit can easily be isolated. Nevertheless, we list these results to $6^{\rm th}$ power in eccentricity, ${\mathcal{O}(e^7)}$, as part of the Supplemental Material~\cite{kp:suppl}.\footnote{This means seventh and higher powers of eccentricity ($e$) have been neglected while writing the results.} We also provide for completeness, the relations connecting the variables of general and elliptical orbits derived from the results presented in \cite{Klein:2010ti, Klein:2018ybm}. Throughout the paper, we only list the spin contributions to both inputs and results expressions to avoid unnecessary duplication of nonspinning results already available in the literature, although, leading nonspinning expressions are included for ease in PN counting. Finally, complete nonspinning contributions to 3PN and spinning contributions to 2PN level for binaries in elliptical orbits to $\mathcal{O}(e^7)$ are listed when writing the expressions to the Supplemental Material~\cite{kp:suppl}.      
\vskip 5pt
Organization of this paper is as follows. In Sec.~\ref{sec:PNinputs}, we recall from literature the necessary PN inputs including -- relations connecting the spherical harmonic modes of gravitational waveforms to a set of \textit{radiative} multipole moments, relations connecting \textit{radiative} and \textit{source} multipole moments as well as necessary expressions for the source moments and equations of motion (EoM). We list our general orbit results (spin contributions to each contributing mode) in terms of a set of radial position and velocity variables in Sec.~\ref{sec:genOrb_hlm}. Next, we specialize to the case of elliptical orbit and list our results for this case in Sec.~\ref{sec:qk} as well as in Appendix \ref{sec:qk_arb} and also as Supplemental Material~\cite{kp:suppl}. Finally, we provide a summary of results and conclude the paper in Sec.~\ref{sec:conclusion}.
\section{Inputs from the post-Newtonian theory}
\label{sec:PNinputs}
\subsection{Spherical harmonic decomposition of gravitational waveforms}
\label{sec:SphHarDec}
Spherical harmonic modes of the gravitational waveform ($h^{\ell m}$) are expressed in terms of the standard (non-STF\footnote{STF designates symmetric and trace-free.}) radiative mass multipoles ($U^{\ell m}$) and radiative current multipoles ($V^{\ell m}$) as \,\citep{Kidder:2007rt, Blanchet:2008je}
\begin{equation}
    h^{\ell m} =- \frac{G}{\sqrt{2}Rc^{\ell+2}}\left[U^{\ell m}-\frac{{ i}}{c}V^{\ell m}\right]\,,
    \label{eq:hlmExp}
\end{equation}
where $G$ is the universal gravitational constant, $R$ is the distance to the source, and $c$ is the speed of light.
\vskip 5pt
It was observed in Ref.~\citep{Kidder:2007rt} and explicitly shown in Ref.~\cite{Faye:2012we}, that, if the binary's orbit is restricted to a plane (as would happen in the absence of spin-precession), then there exists a mode separation. For instance, all ``even" $\ell+m$ modes would depend only on radiative \textit{mass} multipoles ($U^{\ell m}$) and all ``odd" $\ell+m$ modes on \textit{current} radiative multipoles ($V^{\ell m}$). This observation leads to the following simplified expressions for each mode:
\begin{align}
h^{\ell m} &=-\frac{G}{\sqrt{2} R c^{\ell+2}}\times
\begin{cases}
U^{\ell m} & \text { if } \ell+m \text { is even, } \\ \\
-{i\over c} V^{\ell m} & \text { if } \ell+m \text { is odd. } 
\end{cases}
\label{eq:hlmEvenOdd}
\end{align}
\vskip 5pt
The standard (non-STF) moments ($U^{\ell m}$, $V^{\ell m}$) are related to corresponding STF moments ($U_L$, $V_L$) as follows:
\begin{subequations}
\label{eq:UVlmUVL}
\begin{align}
    \label{eq:UlmUL}
    U^{\ell m} &= \frac{4}{\ell!}\sqrt{\frac{(\ell+1)(\ell+2)}{2\ell(\ell-1)}}\alpha^{\ell m}_{L} U_{L}\,,\\
    V^{\ell m} &= -\frac{8}{\ell!}\sqrt{\frac{\ell(\ell+2)}{2(\ell+1)(\ell-1)}}\alpha^{\ell m}_{L} V_{L}\,.
    \label{eq:VlmVL}
\end{align}
\end{subequations}
\vskip 5pt
In the above, $L$=$i_{1}\,i_{2}\cdots i_{\ell}$ represents a multi-index comprising of $\ell$ spatial indices. The object $\alpha^{\ell m}_{L}$ is again a STF tensor that connects the usual basis of spherical harmonics ($Y^{\ell m}$) to another set of STF tensors, $N_{L} = N_{ i_{1}}...N_{i_{\ell}}$ as \cite{Blanchet:2008je}
\begin{equation}
    \alpha_{L}^{\ell m}=\int d \Omega N_{\langle L\rangle} \overline{Y}^{\ell m}\,.
    \label{eq:alphalm}
\end{equation}
\vskip 5pt
Note that $\bm{N}$ is a unit vector pointing toward the detector along the line joining the source and the detector. For instance, if we fix the binary's plane to be the $x$-$y$ plane ${\bm N}$ in terms of angles ($\Theta$, $\Phi$) giving the binary's location reads
\begin{align}
{\bm N}=\sin\Theta\cos\Phi\,\bm{\hat{x}}+\sin\Theta\sin\Phi\,\bm{\hat{y}}+\cos\Theta\,\bm{\hat{z}}\,. 
\label{eq:Neqn}
\end{align}
\vskip 5pt
Finally, the $Y^{\ell m}$ in terms of location angles ($\Theta$, $\Phi$) are given by the following relation:
\begin{align}
    Y^{\ell m}(\Theta, \Phi) &= (-)^{m} \frac{1}{2^{\ell} \ell !}\left[\frac{2 \ell+1}{4 \pi} \frac{(\ell-m) !}{(\ell+m) !}\right]^{1 / 2} \mathrm{e}^{i m \Phi}\, \nonumber \\
    &(\sin \Theta)^{m} \frac{\mathrm{d}^{\ell+m}}{\mathrm{~d}(\cos \Theta)^{\ell+m}}\left(\cos ^{2} \Theta-1\right)^{\ell}\,.
    \label{eq:SphHarExp}
\end{align}
\subsection{Source moments}
\label{sec:sourcemoms}
Using the inputs of Sec.~\ref{sec:SphHarDec} each mode can explicitly be expressed in terms of a mass-type or current-type radiative multipole ($U_L$, $V_L$). 
In the PN-MPM approach ~\cite{Blanchet:2013haa, Blanchet:2000ub, Blanchet:2001aw, Blanchet:2004bb, Blanchet:2003gy, Blanchet:2005tk}, these radiative mass and current multipole moments ($U_L$,$V_L$) are related to a set of  source multipole moments ($I_L$, $J_L$) through a set of mass and current type canonical moments ($M_L$, $S_L$). However, as we shall see below, based on the accuracy of inputs that are required in the current work, the radiative multipoles ($U_L$,$V_L$) appearing in mode expressions can simply be replaced by source moments ($I_L$,$J_L$). At 2PN the two sets of moments are related as\footnote{In principle, there exist hereditary contributions starting at 1.5PN order. However, these can be ignored when computing spin effects to 2PN order.}
\begin{subequations}
\label{eq:RadtoSouMoms}
\begin{align}
    \label{eq:UradToIL}
    U_{L}(U) &= I^{(\ell)}_{L}(U) + \mathcal{O}\bigg(\frac{1}{c^5}\bigg)\,, \\
    V_{L}(U) &= J^{(\ell)}_{L}(U)  + \mathcal{O}\bigg(\frac{1}{c^5}\bigg)\,.
    \label{eq:VradToJL}
\end{align}    
\end{subequations}
\vskip 5pt
In the above, superscript denotes $\ell$th time-derivatives of source moments and $\mathcal{O}(1/c^5)$ indicates that contributions beyond 2PN have been ignored.\footnote{An expression is said to have an accuracy of ${(n/2)}$PN if the results are expressed as a power series in (1/c) to its $n$th power.}
Source multipole moments ($I_L, J_L$) are further expressed as a PN series in terms of binary parameters. In this work we focus on computing the spin effects in $h^{\ell m}$ to 2PN order for systems with nonprecessing spins. Specifically, spin-orbit and spin-spin effects are computed. In order to have sufficient clarity when listing relevant inputs and desired results, we choose to split the two spin contributions. Further, to enable easy PN counting we also provide leading nonspinning (NS) contributions to various quantities. With this view in mind, we express the source multipoles as\footnote{The SS contributions from current moments ($J_L$) appear beyond 2PN order and hence are not included here.}
\begin{subequations}
\label{eq:SouMomsStruc}    
\begin{align}
    \label{eq:ILStruc}
    I_{L} &= I^{{\rm{NS}}}_{L}+I^{{\rm{SO}}}_{L}+I^{{\rm{SS}}}_{L},\,\\
    J_{L} &= J^{{\rm{NS}}}_{L}+J^{{\rm{SO}}}_{L}\,,
    \label{eq:JLStruc}
\end{align}    
\end{subequations}
where NS, SO and SS in the superscript denote the nonspinning, spin-orbit, and spin-spin pieces of the moments. 
\vskip 5 pt
It may be important to note at this stage that, except for the ($\ell=2, |m|=1$) mode, computing all other modes, ($\ell, |m|$)=$((2,2),\, (3,3),\, (3, 2),\, (3, 1),\, (4, 3),\, (4, 1))$ through 2PN order requires only ``leading" inputs (NS, SO, or SS). The $(2, \,1)$ mode, on the other hand, requires  1PN inputs as it would contribute both at 1PN and 2PN level. Further, computing $(2,0)$ and $(3,0)$ modes also requires only leading inputs. It should be clear from Eq.~\eqref{eq:hlmEvenOdd} [when combined with Eq.~\eqref{eq:UVlmUVL} and Eq.~\eqref{eq:RadtoSouMoms}] that the $(2, 1)$ mode depends on the current quadrupole moment ($J_{ij}$) and hence spin contributions to this moment are required with 1PN accuracy [see Eq.~\eqref{eq:JijSO} below]. Computation of all other modes requires source moments with only leading SO and SS effect. Explicit expressions for source moments with each of these contributions have been computed in Refs.~\cite{Kidder:1995zr, Blanchet:2006gy, Arun:2008kb, Buonanno:2012rv}. Below we reproduce these expressions, with the accuracy that is needed in our calculations. These are expressed in terms of position and velocity vectors, ($x^i, v^i$) or ($\bm{x}, \bm{v}$), and spin vectors ($\bm{S}, \bm{\Sigma}$). The spin vectors (${\bm S}$, ${\bm \Sigma}$) are useful combinations of individual spin vectors ($\bm{S_{1,2}}$) and component masses ($m_{1,2}$), where subscript labels represent the two compact objects of the binary. They read
\begin{subequations}
 \label{eq:SpinVar}
\begin{align}
    \bm{S} &\equiv \bm{S}_1 + \bm{S}_2\,,\\ \bm{\Sigma}
    &\equiv M\left[\frac{\bm{S}_2}{m_2} -
    \frac{\bm{S}_1}{m_1}\right]\,,
\end{align}
\end{subequations}
where $M$ is the total mass of the binary.
\vskip 5 pt
\vskip 5pt
The only moment needed with subleading spin corrections is the current quadrupole moment $(J_{ij})$. The relative 1PN SO and leading NS pieces read  
\begin{widetext}
\begin{subequations}
\label{eq:Jij}
\begin{align}
\label{eq:JijNS}
J^{\rm{NS}}_{ij} &= - M\nu\,  \delta \,\varepsilon_{a b\langle i} x^{j \rangle}x^{ a}v^b + \mathcal{O} \left( \frac{1}{c^2}\right)\,, \\  
\label{eq:JijSO}
J^{\rm{SO}}_{ij} &= \frac{\nu}{c}\Bigg(-\frac{3}{2} \,x^{\langle
i}\, \Sigma^{ j\rangle}\Bigg) + \frac{\nu}{c^3}\Bigg\{\frac{3}{7}\left(1-\frac{16\nu}{3}\right)
r \, \dot r\,v^{\langle i}\, \Sigma^{j\rangle} +
\frac{3}{7} \,\delta \, r \, \dot r
\,v^{\langle i}\,S^{j\rangle} +
\bigg[\frac{27}{14}\left(1-\frac{109\nu}{27}\right) (\bm{v}\cdot\bm{\Sigma}) + \frac{27}{14} \delta
\,(\bm{v}\cdot\bm{S})\bigg]\, \nonumber \\
&x^{\langle i}\, v^{j\rangle} +
\left[-\frac{11}{14}\left(1-\frac{47\nu}{11}\right)
(\bm{x}\cdot\bm{\Sigma}) - \frac{11}{14} \delta
\,(\bm{x}\cdot\bm{S})\right]  v^{\langle i} \, v^{j\rangle} +
\bigg[\frac{19}{28}\left(1+\frac{13\nu}{19}\right) \frac{G M}{r}
-\frac{29}{28} \left(1-\frac{143\nu}{29}\right) v^2\bigg] x^{\langle
i} \, \Sigma^{ j\rangle}\, \nonumber \\
&+
\left[-\frac{4}{7}\left(1-\frac{31\nu}{8}\right) (\bm{x}\cdot
\bm{\Sigma}) - \frac{29}{14} \delta
\,(\bm{x}\cdot \bm{S})\right] \frac{G M}{r^3}  \, x^{\langle
i}\, x^{j\rangle} + \left[-\frac{1}{14}\frac{G M}{r} -
\frac{2}{7} v^2\right] \delta\, x^{\langle i}\, S^{j\rangle}\Bigg\} +\mathcal{O}\left(\frac{1}{c^5}\right)\,. \\ \nonumber
\end{align}
\end{subequations}
\end{widetext}
All other moments are needed with only leading spin effects. The NS, SO, and SS pieces for relevant moments read 
\begin{subequations}
\label{eq:spinmultmoms}
\begin{align}
\label{eq:IijSO}
 I^{\rm{NS}}_{ij} &=  M\nu \, x^{\langle i}x^{j\rangle} + \mathcal{O} \left( \frac{1}{c^2}\right)\,, \\
 I^{\rm{SO}}_{ij} &= \frac{8\nu}{3c^{3}}\,x^{\langle i}\left[\bm{v} \times \left(\bm{S}+\delta \bm{\Sigma}\right)\right]^{j \rangle} \nonumber\\&-\frac{4\nu}{3c^{3}}\,v^{\langle i}\left[\bm{x} \times \left(\bm{S}+\delta \bm{\Sigma}\right)\right]^{j \rangle} + \mathcal{O} \left( \frac{1}{c^5}\right)\,, \\
 I^{\rm{SS}}_{ij} &= - \frac{\kappa_1}{m_1 c^4} S_1^{\langle i} S_1^{j\rangle} - \frac{\kappa_2}{m_2 c^4} S_2^{\langle i} S_2^{j\rangle} + \mathcal{O} \left( \frac{1}{c^6}\right)\,, \\ \nonumber \\
\label{eq:IijkSO}
I^{\rm{NS}}_{ijk} &= -M\nu \, \delta  \, x^{\langle i} x^j x^{k \rangle} + \mathcal{O} \left( \frac{1}{c^2}\right)\,, \\  
I^{\rm{SO}}_{ijk} &=  \frac{\nu}{c^3} \, \Bigg[-\frac{9}{2}\,\delta
\,x^{\langle i}x^j \left(\bm{v}\times\bm{S}\right)^{k\rangle}
\nonumber\\&-
\frac{9}{2}\,\left(1-\frac{11\nu}{3}\right)\,x^{\langle i}x^j \left(\bm{v}\times\bm{\Sigma}\right)^{k\rangle} \nonumber\\&+3\,\delta \,x^{\langle i} v^j \left(\bm{x}\times\bm{S}\right)^{k\rangle}+
3\,\left(1-3\nu\right) x^{\langle i}v^j \left(\bm{x}\times \bm{\Sigma}\right)^{k\rangle}\Bigg]\, \nonumber \\
&+\mathcal{O}\left(\frac{1}{c^5}\right) \,,\\ \nonumber  \\
\label{eq:JijkSO}
J^{\rm{SO}}_{ijk} &= \frac{\nu}{c}\Bigg(2 x^{\langle i} x^{j} S^{k\rangle}+2 \delta\,  x^{\langle i} x^{j} \Sigma^{k\rangle}\Bigg)+\mathcal{O}\left(\frac{1}{c^3}\right)\,, \\ \nonumber  \\
\label{eq:JijklSO}
J^{\rm{SO}}_{ijkl} &=
-\frac{5\nu}{2 c} \, \Bigg[\delta\, x^{\langle i}x^j x^k S^{l\rangle}+
\left(1-3\nu\right)x^{\langle i}x^{j} x^{k} \Sigma^{l\rangle}\Bigg]\nonumber\\&+\mathcal{O}\left(\frac{1}{c^3}\right)\,.
\end{align}
\end{subequations}
\vskip 5pt
In the above expressions, $\nu$=$(m_1 m_2/M^2)$ is the symmetric mass ratio parameter and $\delta$=$(m_{1}-m_{2})/M$, with $m_1$, $m_2$ being component masses. (Note that the above definitions lead to $\delta=\pm\sqrt{1-4 \nu}$ depending upon the sign of the difference $m_1-m_2$.) Note also, the angular brackets ($\langle\rangle$) around indices indicate that expressions are symmetric and trace-free in indices they enclose. Additionally, $r=|\bm{x}|$, is the binary's relative separation and $\dot{r}=({\bm{x}\cdot \bm{v})/r}$ the radial component of the relative velocity vector. 
Finally, the parameters $\kappa_{1,2}$ characterize the spin-induced mass quadrupole moment of the two components, defined such that for black holes $\kappa_{1,2}=1$.
\subsection{Equations of motion}
\label{sec:eom}
Equations of motion, which help writing mode expressions completely in terms of position and velocity variables, have been computed with the accuracy desired in \citep{Buonanno:2012rv}. Similar to the source moments these have contributions from SO and SS effects. These are expressed in terms of vectors along the relative separation ($\bm{x}$) and relative velocity ($\bm{v}$) and involve suitable combinations of spin vectors introduced in \citep{Kidder:1995zr, Buonanno:2012rv}. 
We reproduce these here for the convenience of the reader. Again, we split the related expression in terms of NS, SO, and SS pieces for clarity in presentation. The structure for EoM (or simply acceleration, ${\bm{ a}}$) is as follows:
\begin{eqnarray}
\bm{a}&=&
\bm{a}^{{\rm{NS}}} +
\bm{a}^{{\rm{SO}}} + \bm{a}^{{\rm{SS}}}
\,,
\label{eq:accStruc}
\end{eqnarray}
with, 
\begin{subequations}
\begin{align}
\label{eq:accExpNS}
\bm{a}^{{\rm{NS}}} &= -\frac{G M}{r^3}\bm{x}  -\frac{1}{c^2}\frac{G M}{r^3} \Bigg\{\bigg[(1+3\nu)v^2 - \frac{3}{2}\nu\dot r^2 \nonumber \\
&- \frac{4 G M}{r}\left(1+\frac{\nu}{2}\right)\bigg]\bm{x}\, -\,4 r \dot{r} \Big(1-\frac{\nu}{2}\Big) \bm{v}\Bigg\} + \mathcal{O}\left(\frac{1}{c^4}\right),\\
\bm{a}^{{\rm{SO}}} &= \frac{1}{c^3}\frac{G M}{r^3}\left(\frac{G M}{r}\right) \Bigg\{\left[\left(\frac{\bm{x} \times \bm{v}}{r}\right)\cdot \left( \frac{12\bm{S}}{G M^2} \right.\right.\nonumber\\&\left.\left.+\frac{6 \delta \bm{\Sigma}}{G M^2}\right)\right]\bm{x}
-  r\bigg[\bm{v}\times \bigg( \frac{7\bm{S}}{G M^2}+  \frac{3\delta \bm{\Sigma}}{G M^2}\bigg)\bigg] \nonumber\\&+ 3\dot{r} \bigg[\bm{x}\times \bigg( \frac{3\bm{S}}{G M^2} + \frac{\delta \bm{\Sigma}}{G M^2}\bigg)\bigg] \Bigg\} + \mathcal{O}\left(\frac{1}{c^5}\right)\,,
\label{eq:accExpSO}
\end{align}
\vspace{-0.5cm}
\begin{align}
\bm{a}^{{\rm{SS}}} &=-{3\over 2}\frac{1}{c^4}\frac{G M}{r^3}\left(\frac{G M}{r}\right)^2 \Bigg[ \bm{x} \,
\bigg(\frac{\bmSeffp}{G M^2} \cdot \frac{\bmSeffm}{G M^2}\bigg) \, \nonumber \\
& + \bigg(\bm{x} \cdot
\frac{\bmSeffp}{G M^2}\bigg) \frac{\bmSeffm}{G M^2} +  \bigg(\bm{x} \cdot \frac{\bmSeffm}{G M^2}\bigg) \frac{\bmSeffp}{G M^2} \, \nonumber \\
& - \frac{5  \bm{x}}{r^2} \, \bigg(\bm{x} \cdot \frac{\bmSeffp}{G M^2}\bigg) \bigg(\bm{x} \cdot
\frac{\bmSeffm}{G M^2}\bigg) \Bigg] + \mathcal{O}\left(\frac{1}{c^6}\right)\,.
\label{eq:accExpSS}
\end{align}
\end{subequations}
\vskip 5pt
Note again, the NS part above is 1PN accurate. This again is keeping in mind the computation of the ($\ell$=2, $m$=1) mode which has leading (1PN) and subleading (2PN) contributions. For all other modes, contributing to the 2PN waveform Newtonian acceleration in the NS piece together with SO and SS pieces should suffice. Below we summarize notation of \cite{Buonanno:2012rv} used in expressing the EoM.
The set of spin vectors ($\bm{S_0^+}$, $\bm{S_0^{-}}$) that appear in the SS part of the EoM are combinations of individual spin vectors (${\bm S_1}$, ${\bm S_2}$),  component masses ($m_{1,2}$), and parameters characterizing spin-induced effects ($\kappa_{1, 2}$). These read
\begin{subequations}
\label{eq:Seffpm}
\begin{align}
    \bm{S}^{+}_{0} = \frac{M}{m_{1}} \bigg(\frac{\kappa_{1}}{\kappa_{2}}\bigg)^{1/4} \left(1+\sqrt{1-\kappa_{1}\kappa_{2}}\right)^{1/2} {\bm{S_{1}}}\nonumber\\\
    +\frac{M}{m_{2}} \bigg(\frac{\kappa_{2}}{\kappa_{1}}\bigg)^{1/4} \left(1-\sqrt{1-\kappa_{1}\kappa_{2}}\right)^{1/2} {\bm{S_{2}}}\,,\\
    \bm{S}^{-}_{0} = \frac{M}{m_{2}} \bigg(\frac{\kappa_{2}}{\kappa_{1}}\bigg)^{1/4} \left(1+\sqrt{1-\kappa_{1}\kappa_{2}}\right)^{1/2} {\bm{S_{2}}}\nonumber\\\
    +\frac{M}{m_{1}} \bigg(\frac{\kappa_{1}}{\kappa_{2}}\bigg)^{1/4} \left(1-\sqrt{1-\kappa_{1}\kappa_{2}}\right)^{1/2} {\bm{S_{1}}}\,.
\end{align}
\end{subequations}
Additionally, expressions for $\ddot{r}$ and $\dot{v}$ will also be required and can be computed using the following relations,
\begin{subequations}
\begin{align}
\dot{v}&=\frac{\bm{a}\cdot \bm{v}}{v}\,,\\
\ddot{r}&=\frac{1}{r}\left[\left(v^2-\dot{r}^2\right)+{\bm a}\cdot\bm{x}\right]\,.
\label{eq:vdot_rdotdot}
\end{align}
\end{subequations}

with $v=|\bm{v}|$. 
\section{Spherical Harmonic modes with spin effects to 2PN : General orbits, Nonprecessing case}
\label{sec:genOrb_hlm}
We now have all the inputs that are needed to compute spin contributions up to 2PN in spherical harmonic mode expressions. While the computations are algebra intensive they are straightforward.\footnote{All computations are performed using \textit{Mathematica} and the tensor algebra package, xA{\tiny{CT}}~\citep{XAct:xTensor}.} It should be clear that the computation of radiative multipoles ($U_L, V_L$) and hence the computation of modes ($h^{\ell m}$) involves computing time derivatives of source multipoles [see for instance, Eq.~\eqref{eq:RadtoSouMoms}]. Further, as pointed out in Sec.~\ref{sec:eom}, EoM help writing mode expressions solely in terms of position and velocity variables. [With source moments written in terms of position and velocity variables, time derivatives of source moment will induce acceleration terms ($\bm{a}, \ddot{r}, \dot{v}$) and EoM can be used to reexpress the time derivatives again in terms of position and velocity variables.] Results expressed in terms of position and velocity, valid for all possible orbits, are referred to as \emph{general orbit} results. In this section, we list spin contributions through 2PN for each relevant mode in terms of position and velocity variables, namely, the radial separation ($r$), radial velocity ($\dot{r}$), orbital phase ($\phi$), and the angular velocity ($\dot{\phi}$). 
Note also that, from here onward, we specialize to the case of nonprecessing systems. Note however, that inputs of Sec.~\ref{sec:PNinputs} together with evolution equations for spin vectors ($\bm{S}_{1, 2}$) listed in \cite{Buonanno:2012rv} can be used to compute waveforms including the effect of precession. For us, the choice to work with a nonprecessing system drastically simplifies the resulting expressions and helps us provide a completely analytical model, which otherwise would be a semianalytical prescription in the absence of analytical solutions to spin-evolution equations for binaries in noncircular orbits.
\vskip 5 pt
The first obvious simplification occurs due to the use of spin vectors of constant magnitude ($\bm{S}^c_{1, 2}$ with $\bm{S}^c_{i}\cdot\bm{\dot S}^c_{i}=0$) as opposed to the usual spin vectors ($\bm{S}_{1, 2}$) as in \cite{Blanchet:2006gy, Buonanno:2012rv}. Further, unit vectors ($\bm{\hat n}$, $\bm{\hat \lambda}$) along the relative separation and velocity vector ($\bm{x}$, $\bm{v}$), together with the unit vector along the orbital angular momentum, $\bm{\hat{\ell}}$, form an orthonormal triad.\footnote{Note that we use the same letter, $\ell$, for representing the mode number in $h^{\ell m}$ and the orbital angular momentum unit vector, however, throughout the paper we express the angular momentum unit vector as $\bm{\hat{\ell}}$.} This further helps in expressing results solely in terms of $(\bm{\hat{\ell}}\cdot\bm{S}_{\rm{c}})$ and $(\bm{\hat{\ell}}\cdot\bm{\Sigma}_{\rm{c}})$, where $\bm{S_c}$ and $\bm{\Sigma_c}$ in terms of spin vectors ($\bm{S}$, $\bm{\Sigma}$) are defined in \cite{Buonanno:2012rv} and we simply reproduce them here for the ease of the reader,\footnote{See Sec. IV E of \cite{Buonanno:2012rv} for a detailed discussion.}
\begin{subequations}
\label{eq:spinconstScSigmac}
\begin{align}
\label{eq:spinconstSc}
\bm{S}_{\rm c} &= \bm{S}+\frac{G M \nu}{r c^2}\Big[2 \bm{S}+
\delta\,\bm{\Sigma}\Big] + \mathcal{O}\left(\frac{1}{c^3}\right)\,, \\
\label{eq:spinconstSigmac}
\bm{\Sigma}_{\rm c}&= \bm{\Sigma}+\frac{G M}{r c^2}\Big[\delta\,\bm{S}
+(1-2\nu)\bm{\Sigma}\Big] + \mathcal{O}\left(\frac{1}{c^3}\right). 
\end{align}
\end{subequations}
\vskip 5pt
We are now ready to write final expressions for spin effects in each mode contributing to gravitational waveforms through 2PN order. Following \cite{Mishra:2016whh} we work with a source frame similar to the one chosen for spinning systems in Refs.~\cite{Arun:2008kb, Buonanno:2012rv}. As was pointed out in \cite{Mishra:2016whh}, this choice of the source frame is different from the one in \cite{Blanchet:2008je} which does not discuss spin contributions. Nevertheless, mode expressions in conventions used in \cite{Blanchet:2008je} can be obtained by simply dividing by a factor of $i^m$ to our results below.  [This is also the reason we have introduced a multiplicative factor of $i^{2m}$ when presenting the results; see for instance, Eq.~(\ref{eq:genOrb_hlm}) below.) In terms of generic position and velocity variables ($r, \dot{r}, \phi, \dot{\phi}$), mode expressions take the following form,
\begin{equation}
    h^{\ell m} = -4\,{i^{2m}}\sqrt{\frac{\pi}{5}}\,\left(\frac{G M \nu}{c^{4} R}\right) \mathrm{e}^{-i m \phi}H^{\ell m}
\label{eq:genOrb_hlm}
\end{equation}
with
\begin{align}
    H^{\ell m}=H^{\ell m}_{\rm NS}+H^{\ell m}_{\rm SO}+H^{\ell m}_{\rm SS}\,,
    \label{eq:Hlmstruct}
\end{align}
where NS, SO, and SS contributions for all relevant modes read
\begin{widetext}
\begin{subequations}
\label{eq:genOrb_Hlm}
\begin{align}
    H^{22}_{\rm{NS}} &= \frac{G M}{r}+\left(i \dot{r}+r \dot{\phi }\right)^2 +\mathcal{O}\left(\frac{1}{c^2}\right) \,, \\
    H^{22}_{\rm{SO}} &= -\frac{1}{3}\frac{1}{c^3}\left(\frac{G M}{r}\right)^{2} \bigg[\left(4 i \dot{r}+7 r \dot{\phi }\right) \left(\bm{\hat{\ell}} \cdot \bm{s}_{c}\right)+\delta 
   \left(-2 i \dot{r}+r \dot{\phi }\right) \left(\bm{\hat{\ell}} \cdot \bm{\sigma}_{c}\right)\bigg] +\mathcal{O}\left(\frac{1}{c^5}\right)\,, \\
   H^{22}_{\rm{SS}} &= \frac{3}{2}\frac{1}{c^4}\left(\frac{G M}{r}\right)^{3} \bigg[\left(\bm{s}_{0}^{+}\cdot\bm{s}_{0}^{-}\right)\bigg] +\mathcal{O}\left(\frac{1}{c^5}\right)\,, \\
   H^{21}_{\rm{NS}} &= -\frac{2}{3}\frac{1}{c} \left(\frac{G M}{ \,r}\right) \delta \left(r   \dot{\phi }\right) + \mathcal{O}\left(\frac{1}{c^3}\right) \,, \\
   H^{21}_{\rm{SO}} &= -\frac{1}{c^2}\left(\frac{G M}{ r}\right)^{2} \left(\bm{\hat{\ell}} \cdot \bm{\sigma}_{c}\right) + \frac{1}{42}\frac{1}{c^4}\left(\frac{G M}{
    r}\right)^{2} \Bigg\{2\delta  \bigg[ \frac{24 G M}{r} + 2\dot{r}^{2} + 48 i \dot{r} \left(r \dot{\phi}\right) + 62   \left(r \dot{\phi}\right)^{2}
    \bigg] \left(\bm{\hat{\ell}} \cdot \bm{s}_{c}\right) - \bigg[-\frac{202 G M}{r}
   \, \nonumber \\
   & \times \left(1-\frac{37 \nu }{101}\right) - 109
   \left(1-\frac{68 \nu }{109}\right)  \dot{r}^{2} + 198 i \dot{r} \left(r\dot{\phi}\right) \left(1+\frac{109
   \nu }{99}\right)   - 40 \left(r \dot{\phi}\right)^{2} \left(1-\frac{29\nu}{5}\right)   
   \bigg] \left(\bm{\hat{\ell}} \cdot \bm{\sigma}_{c}\right) \Bigg\}+\mathcal{O}\left(\frac{1}{c^5}\right)\,, \\
   H^{20}_{\rm{NS}} &=  \sqrt{\frac{2}{3}}\Bigg\{\frac{G M}{r}- \left(r \dot{\phi}\right)^{2}-\dot{r}^{2}\Bigg\}+\mathcal{O}\left(\frac{1}{c^2}\right)\,, \\
    H^{20}_{\rm{SO}} &=  -\sqrt{\frac{2}{3}}\frac{1}{c^3}\left(\frac{G M}{r}\right)^{2}  \left(r\dot{\phi}\right) \Bigg\{ 5 \left(\bm{\hat{\ell}} \cdot \bm{s}_{c}\right)+ 3 \delta \left(\bm{\hat{\ell}} \cdot \bm{\sigma}_{c}\right)\Bigg\}+\mathcal{O}\left(\frac{1}{c^5}\right)\,, \\
   \label{eq:20SSGen}
     H^{20}_{\rm{SS}} &=  \sqrt{\frac{3}{2}}\frac{1}{c^4}\left(\frac{ G M}{ r}\right)^{3} \bigg[\left(\bm{s}_{0}^{+}\cdot\bm{s}_{0}^{-}\right)\bigg]+\mathcal{O}\left(\frac{1}{c^5}\right)\,, \\
   H^{33}_{\rm{NS}} &= -\frac{\sqrt{5}}{2 \sqrt{42}}\frac{1 }{c}\delta \,  \Bigg\{2 \left(i \dot{r}+r \dot{\phi
   }\right)^3+\frac{G M}{r} \left[4 i \dot{r}+7 r \dot{\phi }\right]\Bigg\}  +\mathcal{O}\left(\frac{1}{c^3}\right)\,, \\
   H^{33}_{\rm{SO}} &=\frac{\sqrt{5}}{4 \sqrt{42}}\frac{1}{c^4} \left(\frac{ G M }{\, r}\right)^{2} \Bigg\{\delta\,  \left[-10 \dot{r}^2+38 i
   \dot{r} \left(r\dot{\phi }\right)+43 \left(r \dot{\phi }\right)^2\right] \left(\bm{\hat{\ell}} \cdot \bm{s}_{c}\right)+\bigg[-\frac{4 G M
   }{r}\left(1-5 \nu \right)+2 \dot{r}^2 (1-5 \nu ) +2 i \dot{r} \, \nonumber \\ 
   & \times \left(r\dot{\phi
   }\right) \left(1+\nu \right) + 19 \left(r \dot{\phi }\right)^2 \left(1-\frac{53\nu}{19}  \right) \bigg] \left(\bm{\hat{\ell}} \cdot \bm{\sigma}_{c}\right)\Bigg\} +\mathcal{O}\left(\frac{1}{c^5}\right)\,, \\
   H^{32}_{\rm{NS}} &= \frac{\sqrt{5}}{6\sqrt{7}}\frac{1}{c^2}\left(\frac{G M}{ r}\right) \left(1-3 \nu \right) \left(r\dot{\phi }\right) \bigg[i
   \dot{r}+4 r \dot{\phi }\bigg] +\mathcal{O}\left(\frac{1}{c^4}\right)\,, \\
   H^{32}_{\rm{SO}} &= \frac{\sqrt{5}}{3 \sqrt{7}}\frac{1}{c^3}\left(\frac{ G M}{r }\right)^{2} \bigg[i \dot{r}+4 r \dot{\phi }\bigg]
   \bigg[\left(\bm{\hat{\ell}} \cdot \bm{s}_{c}\right)+\delta  \left(\bm{\hat{\ell}} \cdot \bm{\sigma}_{c}\right)\bigg] +\mathcal{O}\left(\frac{1}{c^5}\right)\,, \\
   H^{31}_{\rm{NS}} &= \frac{1}{6 \sqrt{14}}\frac{1}{c}\delta  \Bigg\{6 \left(-i \dot{r}+r \dot{\phi }\right) \left(i \dot{r}+r
   \dot{\phi }\right)^2-\frac{G M}{r} \bigg[12 i \dot{r}+7 r \dot{\phi
   }\bigg]\Bigg\} +\mathcal{O}\left(\frac{1}{c^3}\right)\,, \\
   H^{31}_{\rm{SO}} &= \frac{1}{12 \sqrt{14}}\frac{1}{c^4}\left(\frac{G M}{  r}\right)^{2} \Bigg\{\delta  \left[-10 \dot{r}^2+66 i  \dot{r} \left(r\dot{\phi }\right)+ 59
   \left(r \dot{\phi }\right)^2\right] \left(\bm{\hat{\ell}} \cdot \bm{s}_{c}\right)+\bigg[-\frac{ 4 G M }{r}\left(1-5 \nu \right)+2 \dot{r}^2 \left(1-5 \nu
   \right) + 54 i  \dot{r} \, \nonumber \\ 
   & \times \left(r\dot{\phi }\right) \left(1-\frac{101\nu}{27}  \right) +35 \left(r \dot{\phi }\right)^2 \left(1-\frac{149\nu}{35}  \right)
   \bigg] \left(\bm{\hat{\ell}} \cdot \bm{\sigma}_{c}\right)\Bigg\} +\mathcal{O}\left(\frac{1}{c^5}\right)\,, \\
    H^{30}_{\rm{NS}} &=  \frac{i}{\sqrt{42}}\frac{1}{ c^2} \left(1-3 \nu \right) \frac{G M}{r} \dot{r} \left(r\dot{\phi}\right)+\mathcal{O}\left(\frac{1}{c^4}\right)\,, \\
    H^{30}_{\rm{SO}} &=  i \sqrt{\frac{2}{21}}\frac{1}{c^3}\left(\frac{ G M}{r}\right)^{2} \dot{r}  \bigg[\left(\bm{\hat{\ell}} \cdot \bm{s}_{c}\right)+\delta  \left(\bm{\hat{\ell}} \cdot \bm{\sigma}_{c}\right)\bigg]+\mathcal{O}\left(\frac{1}{c^5}\right)\,, \\
   H^{43}_{\rm{NS}} &= -\frac{2}{3 \sqrt{70}}\frac{1}{c^3}\left(\frac{G M   }{\,r }\right)\delta \left(1-2 \nu \right) \left(r\dot{\phi }\right) \bigg[\frac{ G M}{r}-\frac{1}{2} \dot{r}^2+ \frac{5}{2}
   i  \dot{r} \left(r\dot{\phi }\right)+\frac{23}{4} \left(r \dot{\phi }\right)^2\bigg] +\mathcal{O}\left(\frac{1}{c^5}\right)\,, \\
   H^{43}_{\rm{SO}} &= - \frac{\sqrt{5}}{3 \sqrt{14}}\frac{1}{c^4}\left(\frac{ G M}{ r}\right)^{2} \left[\frac{ G M}{r}-\frac{1}{2} \dot{r}^2+\frac{5}{2} i 
   \dot{r} \left(r\dot{\phi }\right)+ \frac{23}{4} \left(r \dot{\phi }\right)^2\right] \bigg[\delta  \left(\bm{\hat{\ell}} \cdot \bm{s}_{c}\right) + \left(1-3
   \nu \right) \left(\bm{\hat{\ell}} \cdot \bm{\sigma}_{c}\right)\bigg] +\mathcal{O}\left(\frac{1}{c^5}\right)\,, \\
   H^{41}_{\rm{NS}} &= -\frac{2}{7 \sqrt{70}}\frac{1}{c^3}\left(\frac{G M}{r}\right) \delta \left(1-2 \nu \right) \left(r\dot{\phi }\right) \left[ \frac{ G M}{r}-\frac{1}{2}
   \dot{r}^2+\frac{5}{6} i  \dot{r} \left(r\dot{\phi }\right)-\frac{11}{12} \left(r \dot{\phi }\right)^2\right] +\mathcal{O}\left(\frac{1}{c^5}\right)\,, \\
   H^{41}_{\rm{SO}} &=-\frac{\sqrt{5}}{7 \sqrt{2}}\frac{1}{c^4} \left(\frac{ G M}{ r}\right)^{2} \left[\frac{G M}{r}-{1\over 2}\dot{r}^2+{5\over6} i 
   \dot{r} \left(r\dot{\phi }\right)-{11\over12} \left(r \dot{\phi }\right)^2\right] \bigg[\delta  \left(\bm{\hat{\ell}} \cdot \bm{s}_{c}\right)+\left(1-3
   \nu \right) \left(\bm{\hat{\ell}} \cdot \bm{\sigma}_{c}\right)\bigg] +\mathcal{O}\left(\frac{1}{c^5}\right)\,,
\end{align}
\end{subequations}
\end{widetext}
where we have used $v^{2}=\dot{r}^{2}+r^{2}\dot{\phi}^2$. 
Note also in the above, we have redefined the spin variables ($\bm{S}_c, \bm{\Sigma}_c, \bm{S}^{+}_{0}, \bm{S}^{-}_{0}$) as 
\begin{align}
    \bm{s}_{c}&=\frac{\bm{S}_{c}}{G M^2},\,
    \bm{\sigma}_{c}=\frac{\bm{\Sigma}_{c}}{G M^2}\,,  \bm{s}^{+}_{0}=\frac{\bm{S}^{+}_{0}}{G M^2},\, 
    \bm{s}^{-}_{0}=\frac{\bm{S}^{-}_{0}}{G M^2}\,.
    \label{eq:spinVarRedefine}
\end{align}
Following \cite{Mishra:2016whh}, we also express below the spin contributions in terms of symmetric and antisymmetric combinations of dimensionless component spin parameters, $\bm{\chi}^{c}_{i}={\bm{S}^{c}_{i}}/G\,m^{2}_{i}$, useful in data-analysis applications. [Note that this change of spin variables does not affect the leading nonspinning part included in Eq~.\eqref{eq:genOrb_Hlm} and hence is not included in Eq.~\eqref{eq:genOrb_Hlm_chia_chis} to avoid duplication.] With $\bm{\chi}^{c}_{s}=\frac{1}{2}(\bm{\chi}^{c}_{1}+\bm{\chi}^{c}_{2})$ and $\bm{\chi}^{c}_{a}=\frac{1}{2}(\bm{\chi}^{c}_{1}-\bm{\chi}^{c}_{2})$, spin contributions to each mode expressed in terms of ($\bm{\chi}^{c}_{s}$,  $\bm{\chi}^{c}_{a}$) read
\begin{widetext}
\begin{subequations}
\label{eq:genOrb_Hlm_chia_chis}
\begin{align}
H^{22}_{\rm{SO}} &= -\frac{2}{3}\frac{1}{c^3}\left(\frac{ G M}{  r}\right)^{2} \Bigg\{3 \delta  \left(\bm{\hat{\ell}} \cdot \bm{\chi}^{c}_{a}\right) \left(r \dot{\phi }+ i
   \dot{r}\right)+\left(\bm{\hat{\ell}} \cdot \bm{\chi}^{c}_{s}\right) \left[3\left(1-\frac{5 \nu}{3} \right) r \dot{\phi
   } + 3i \dot{r} \left(1-\frac{8 \nu}{3} \right) \right]\Bigg\} +\mathcal{O}\left(\frac{1}{c^5}\right)\,, \\
H^{22}_{\rm{SS}} &= -\frac{3}{2}\frac{1}{c^4}\left(\frac{ G M}{  r}\right)^{3} \Bigg\{2\, \left(\bm{\hat{\ell}} \cdot \bm{\chi}^{c}_{a}\right) \left(\bm{\hat{\ell}} \cdot \bm{\chi}^{c}_{s}\right) \bigg[-(1-2 \nu ) \kappa
   _a-\delta  \kappa _s\bigg]+\left(\bm{\hat{\ell}} \cdot \bm{\chi}^{c}_{s}\right){}^2 \bigg[-2 \nu
   -\delta  \kappa _a-(1-2 \nu ) \kappa _s\bigg] \, \nonumber \\
   &+\left(\bm{\hat{\ell}} \cdot \bm{\chi}^{c}_{a}\right){}^2 \bigg[2 \nu -\delta  \kappa _a -(1-2 \nu ) \kappa_s\bigg]\Bigg\} +\mathcal{O}\left(\frac{1}{c^5}\right)\,, \\
H^{21}_{\rm{SO}} &=  \frac{1}{c^2}\left(\frac{ G M}{r }\right)^{2} \bigg[\left(\bm{\hat{\ell}} \cdot \bm{\chi}^{c}_{a}\right)+\delta\, \left(\bm{\hat{\ell}} \cdot \bm{\chi}^{c}_{s}\right) \bigg] - \frac{1}{42}\frac{1}{c^4}\left(\frac{G M}{ r}\right)^{2} \Bigg\{\delta \left(\bm{\hat{\ell}} \cdot \bm{\chi}^{c}_{s}\right) \Bigg[\frac{154 G M}{r}  \left(1+\frac{\nu }{7}\right)-84 \left(1-\frac{4 \nu }{21}\right) \left(r\dot{\phi}\right)^{2} \, \nonumber \\
&-294 i  \left(1+\frac{13 \nu
   }{147}\right) \dot{r} \left(r\dot{\phi}\right) +105 \dot{r}^{2}   \left(1-\frac{4 \nu }{7}\right)\Bigg]+ \left(\bm{\hat{\ell}} \cdot \bm{\chi}^{c}_{a}\right)
   \Bigg[\frac{154 G M}{r} \left(1+ \frac{59 \nu }{77}\right)-84  \left(1-\frac{22 \nu }{7}\right) \left(r\dot{\phi}\right)^2 \, \nonumber \\
   & -294 i \left(1-\frac{83 \nu }{147}\right) \dot{r} \left(r \dot{\phi} \right)+105 \dot{r}^{2} \left(1-\frac{52 \nu
   }{105}\right)\Bigg]\Bigg\}+\mathcal{O}\left(\frac{1}{c^5}\right)\,, \\
 H^{20}_{\rm{SO}} &=  -2 \sqrt{\frac{2}{3}}\frac{1}{c^3}\left(\frac{ G M}{r}\right)^{2} \left(r\dot{\phi}\right) \bigg[\delta \left(\bm{\hat{\ell}} \cdot \bm{\chi}^{c}_{a}\right) +(1+\nu) \left(\bm{\hat{\ell}} \cdot \bm{\chi}^{c}_{s}\right)\bigg]+\mathcal{O}\left(\frac{1}{c^5}\right)\,, \\
\label{eq:20SSGen_chi}
 H^{20}_{\rm{SS}} &=  \sqrt{\frac{3}{2}}\frac{1}{c^4}\left(\frac{ G M}{r}\right)^{3} \Bigg\{\left(\bm{\hat{\ell}} \cdot \bm{\chi}^{c}_{a}\right)^{2} \bigg[-2 \nu + \delta \kappa_{a} + \kappa_{s}\left(1-2\nu\right)\bigg]+2 \left(\bm{\hat{\ell}} \cdot \bm{\chi}^{c}_{a}\right)\left(\bm{\hat{\ell}} \cdot \bm{\chi}^{c}_{s}\right) \bigg[\delta  \kappa_{s} + \kappa_{a}\left(1-2\nu\right)\bigg] \, \nonumber \\
& + \left(\bm{\hat{\ell}} \cdot \bm{\chi}^{c}_{s}\right)^{2} \bigg[2 \nu+\delta  \kappa_{a} + \kappa_{s} \left(1-2\nu\right) \bigg]\Bigg\}+\mathcal{O}\left(\frac{1}{c^5}\right)\,,\\
H^{33}_{\rm{SO}} &= -\frac{\sqrt{5}}{4 \sqrt{42}}\frac{1}{c^4}\left(\frac{ G M}{ r}\right)^{2} \Bigg\{\delta  \left(\bm{\hat{\ell}} \cdot \bm{\chi}^{c}_{s}\right)
   \bigg[-\frac{ 4G M }{r}\left(1-5 \nu \right)-24 \left(1-\frac{11 \nu }{8}\right)
   \left(r \dot{\phi }\right)^2-36 i \left(1-\frac{13 \nu }{6}\right)
   \dot{r}\left(r \dot{\phi }\right) +12 \dot{r}^2 \, \nonumber \\
   & \times \left(1-\frac{5 \nu }{2}\right)
   \bigg] +\left(\bm{\hat{\ell}} \cdot \bm{\chi}^{c}_{a}\right) \bigg[-\frac{4 G M }{r}\left(1-5 \nu \right)-24
   \left(1-\frac{119 \nu }{24}\right) \left(r \dot{\phi }\right)^2-36 i
   \left(1-\frac{77 \nu }{18}\right) \dot{r}\left(r \dot{\phi }\right)
   +12\dot{r}^2  \, \nonumber \\
   & \times \left(1-\frac{25 \nu }{6}\right)
   \bigg]\Bigg\}+\mathcal{O}\left(\frac{1}{c^5}\right)\,, \\
H^{32}_{\rm{SO}} &= \frac{2 \sqrt{5}}{3 \sqrt{7}}\frac{1}{c^3}\left(\frac{  G M }{ r}\right)^2 \nu \left(\bm{\hat{\ell}} \cdot \bm{\chi}^{c}_{s}\right) \bigg[4 r
   \dot{\phi } + i \dot{r}\bigg] +\mathcal{O}\left(\frac{1}{c^5}\right) \,, \\
H^{31}_{\rm{SO}} &= \frac{1}{12 \sqrt{14}}\frac{1}{c^4}\left(\frac{G M}{r}\right)^{2} \Bigg\{\delta  \left(\bm{\hat{\ell}} \cdot \bm{\chi}^{c}_{s}\right) \bigg[\frac{4 G M }{r}\left(1-5 \nu
   \right)+24 \left(1+\frac{31 \nu }{24}\right) \left(r \dot{\phi
   }\right)^2+12 i \left(1+\frac{35 \nu }{6}\right) \dot{r} \left(r \dot{\phi
   }\right) -12 \dot{r}^2 \, \nonumber \\
   & \times \left(1-\frac{5 \nu }{2}\right)
   \bigg] +\left(\bm{\hat{\ell}} \cdot \bm{\chi}^{c}_{a}\right) \left[\frac{4 G M }{r}\left(1-5 \nu \right)+24
   \left(1-\frac{29 \nu }{8}\right) \left(r \dot{\phi }\right)^2+12 i
   \left(1-\frac{31 \nu }{6}\right) \dot{r}\left(r \dot{\phi }\right)
   -12 \dot{r}^2 \left(1-\frac{25 \nu }{6}\right)
   \right]\Bigg\} \, \nonumber \\
   &+\mathcal{O}\left(\frac{1}{c^5}\right)\,, \\
H^{30}_{\rm{SO}} &= 2 i \sqrt{\frac{2}{21}}\frac{1}{c^3}\left(\frac{G M}{r}\right)^2 \nu \left(\bm{\hat{\ell}} \cdot \bm{\chi}^{c}_{s}\right) \dot{r}+\mathcal{O}\left(\frac{1}{c^5}\right)\,, \\   
H^{43}_{\rm{SO}} &=\frac{\sqrt{5}}{3 \sqrt{14}}\frac{1}{c^4}\left(\frac{ G M  }{ r}\right)^2 \nu \bigg[\left(\bm{\hat{\ell}} \cdot \bm{\chi}^{c}_{a}\right)-\delta  \left(\bm{\hat{\ell}} \cdot \bm{\chi}^{c}_{s}\right)\bigg] \left[\frac{ G M}{r}+ \frac{23}{4} \left(r \dot{\phi }\right)^2+ \frac{5}{2}
   i \dot{r} \left(r \dot{\phi }\right) - \frac{1}{2} \dot{r}^2\right] +\mathcal{O}\left(\frac{1}{c^5}\right)\,, \\
H^{41}_{\rm{SO}} &=\frac{\sqrt{5}}{7 \sqrt{2}}\frac{1}{c^4}\left(\frac{ G M }{ r}\right)^2 \nu  \bigg[\left(\bm{\hat{\ell}} \cdot \bm{\chi}^{c}_{a}\right)-\delta  \left(\bm{\hat{\ell}} \cdot \bm{\chi}^{c}_{s}\right)\bigg] \left[\frac{ G M}{r}-\frac{11}{12} \left(r \dot{\phi }\right)^2+\frac{5}{6}
   i \dot{r} \left(r \dot{\phi }\right) -\frac{1}{2} \dot{r}^2\right] +\mathcal{O}\left(\frac{1}{c^5}\right)\,,
\end{align}   
\end{subequations}
\end{widetext}
where we have used
$\kappa_{s}=(\kappa_{1}+\kappa_{2})/2$ and $\kappa_{a}=(\kappa_{1}-\kappa_{2})/2$.\\

Recently we learned about a related work~\citep{Khalil:2021txt} which explicitly writes general orbital expressions for various modes with spins up to 2PN order in harmonic coordinates and in terms of dimensionless spin variables (${\bm{\chi}_{1}}, {\bm{\chi}_{2}}$) associated with \emph{nonconstant} spins ($\bm{S}_{1},\bm{S}_{2}$) as $\bm{\chi}_{i}={\bm{S}_{i}}/G\,m^{2}_{i}$. [See Eq.~(C1) there for complete expression.] The connection between (${\bm{\chi}_{1}}, {\bm{\chi}_{2}}$) and the \emph{constant} spin variables (${\bm{\chi}^c_{1}}, {\bm{\chi}^c_{2}}$) used above can be obtained by employing Eq.~\eqref{eq:spinconstScSigmac} and they read
\begin{subequations}
\begin{align}
\bm{\chi}^{c}_{1} = \bm{\chi}_{1}+\frac{1}{2}\left(\frac{G M }{c^2 r}\right)\left(1-\delta\right)\bm{\chi}_{1},\\ \bm{\chi}^{c}_{2} = \bm{\chi}_{2}+\frac{1}{2}\left(\frac{G M }{c^2 r}\right)\left(1+\delta\right)\bm{\chi}_{2}. 
\end{align}
\end{subequations}
Using the above connection one can show the correspondence between the results of Ref.~\citep{Khalil:2021txt} and those presented here. We confirm that our results are in complete agreement with those of Ref.~\citep{Khalil:2021txt}.\footnote{Note also that the spin quadrupole constants ($C_{1ES^2}, C_{2ES^2}$) in our notation are $\kappa_{1, 2}$.}
\section{quasi-keplerian representation}
\label{sec:qk}
The general orbit expressions with spin contributions to 2PN in all the relevant spherical harmonic modes were listed in the previous section. Leading expressions for the nonspinning part was also included for ease in PN counting. These results can be applied to compact binaries with nonprecessing spins and in orbits of arbitrary shapes. In this section, we specialize to the case of elliptical orbits by making use of generalized quasi-Keplerian (QK) representation for the conservative dynamics of the spinning compact binaries in elliptical orbits to 2PN \,\citep{Klein:2010ti, Klein:2018ybm}. (Reference \cite{Klein:2018ybm} also includes the corresponding 3PN QK representation of \cite{AIHPA_1985__43_1_107_0,Damour:1988mr, 1993PhLA..174..196S, 1995CQGra..12..983W, Memmesheimer:2004cv} for nonspinning systems in harmonic coordinates.) It may be worth noting that the QK representation presented in \cite{Klein:2018ybm} is in harmonic coordinates or more specifically in \textit{modified} harmonic (MH) coordinates that were introduced to remove the presence of logarithmic terms in expressions for mass quadrupole moment at 3PN order \cite{Memmesheimer:2004cv, Arun:2007sg} and have been extensively used in numerous efforts focusing on modeling compact binary inspirals in elliptical orbits \cite{Mishra:2015bqa, Boetzel:2019nfw, Moore:2016qxz, Boetzel:2017zza, Klein:2010ti, Klein:2018ybm}.
\vskip 5pt
The QK representation involves connecting variables describing compact binaries in general orbits to those characterizing the orbit one wishes to specialize to. Once these relations are established, they can be used to 
convert the general orbit results to corresponding ones for an orbit of interest. The QK representation of \cite{Klein:2018ybm} explicitly writes these relations for spinning systems in an elliptical orbit. Moreover, as far as we know this is the only prescription available that describes the binary's conservative dynamics and evolution of orbital elements to 2PN order for spinning systems and we choose to use these results here. We also keep the symbols and notations as used in \cite{ Klein:2018ybm} to avoid any confusion unless it conflicts with our presentation; in that case, we inform the reader about the differences in  notations adopted in the current work. However, in order to avoid duplication, we only wish to reproduce representation of \cite{Klein:2018ybm} partly, although the results included here should be sufficient to convert the general orbit results of Sec.~\ref{sec:genOrb_hlm} to the elliptical case.
\vskip 5pt
Here, we wish to list expressions for each mode included in Sec.~\ref{sec:genOrb_hlm} specialized to the case of elliptical orbits. Since the general orbit results are explicitly written in terms of a set of radial ($r$,\,$\dot{r}$) and angular variables ($\phi$, and $\dot{\phi}$), converting the general orbit results simply requires using expressions relating these general orbit variables to those of elliptical orbit. Below we reproduce the relations that connect $r$,\,$\dot{r}$\,$\phi$, and $\dot{\phi}$ to a set of variables characterizing an elliptical orbit. They read
\begin{subequations}
\label{eq:qK_r_rdot_phi_phidot}
\begin{align}
    r &= a (1- e_r \cos u) + f_{r}(V)\,,\\
    \dot{r} &= \frac{a \, e_r\, n\, \sin u}{1- e_t \cos u}\,, \\
    \phi &= (1+k)V +f_{\phi}(V)\,, \\
    \dot{\phi} &= \frac{\sqrt{1- e_{\phi}^2}\, (1+k)\, n}{(1- e_t \cos u) (1- e_{\phi} \cos u)}\,.
\end{align}
\end{subequations}
Here, $a$ and $e$ denote the semimajor axis and orbital eccentricity. Note that parameter subscripts ($r, \phi, t$) indicate that the concerned parameter is associated with a specific coordinate. For instance, all three types of eccentricities ($e_{r},\,e_{\phi},\,e_{t}$) are featured in the above relations. (The subscript is dropped in case there is no need to distinguish between them.) Further, the symbols ($n, \,k, \,u$ and $V$),  represent mean motion, advance of periastron, eccentric anomaly, and true anomaly,\footnote{Reference~\cite{Klein:2018ybm} uses the symbol $v$ for true anomaly; although, here we already use $v$ for the magnitude of the velocity vector and thus alternatively use symbol $V$ following \cite{Arun:2007sg} to represent true anomaly.} respectively. Finally, $f_t,\,f_r$ and $f_\phi$ are 2PN corrections to these relations; however, they contribute to spin effects beyond 2PN for \textit{nonprecessing} systems and hence are ignored here. Explicit expressions for QK parameters characterizing the binary's motion in an elliptical orbit in terms of a PN parameter ($y$), time eccentricity ($e_t$) and two spin related parameters ($\beta$, $\gamma_1$), with the accuracy desired in our work, have been listed in Appendix B of Ref.~\cite{Klein:2018ybm} and we do not wish to reproduce them here. Nevertheless, we provide explicit relations for radial and angular position and velocity variables in terms of the parameters chosen in 
\cite{Klein:2018ybm} in Appendix~\ref{sec:qk_orbdyn}. These could directly be used in general orbital expressions for modes presented in Sec.~\ref{sec:genOrb_hlm} in order to reexpress them for the elliptical case.
\vskip 5pt
\noindent
Leading NS and SO, SS contributions up to 2PN to all relevant modes valid for compact binaries in orbits of \textit{arbitrary eccentricity} have been listed in Appendix~\ref{sec:qk_arb}. These are probably the most useful results of the current paper, although,  they have been pushed to the appendix to avoid disruption of the flow of the paper. Instead, in this section we list these contributions as leading eccentric corrections to the circular case by expanding our arbitrary eccentricity results by treating eccentricity as a small parameter. Only leading corrections are included so as to have clarity in presentation and again to maintain the flow of the paper. However, the arbitrary eccentricity results of Appendix~\ref{sec:qk_arb} can be expanded to an arbitrary power of eccentricity and thus become suitable for small to moderate eccentricities. In fact, we do provide these results to $\mathcal{O}(e^7)$ as Supplemental Material~\cite{kp:suppl}. These complement the nonspinning results available to $\mathcal{O}(e^7)$ in eccentricity up to 3PN order and were computed in \cite{Boetzel:2019nfw, Ebersold:2019kdc}. 
\vskip 5pt
Expressions for each relevant GW mode with leading NS and SO, SS terms to 2PN and with leading correction of eccentricity to the circular case read
\begin{widetext}
\begin{equation}
    h^{\ell m} = - i^{2 m}\, \frac{8\,G\,M\,\nu\,x}{c^{2}\,R}\,\sqrt{\frac{\pi}{5}}\,\mathrm{e}^{-im\phi}\,H^{\ell m}
    \label{eq:hlm_qK_leading_e}
\end{equation}
with
\begin{align}
    H^{\ell m}=H^{\ell m}_{\rm NS}+H^{\ell m}_{\rm SO}+H^{\ell m}_{\rm SS}\,,
    \label{eq:Hlmstruct}
\end{align}
where NS, SO, and SS pieces for all relevant modes read
\begin{subequations}
\label{eq:Hlm_qK_leading_e}
\begin{align}
H^{22}_{{\rm{NS}}} &=  \Bigg[1+e \left(\frac{1}{4}\mathrm{e}^{-i l}+\frac{5}{4} \mathrm{e}^{i l}\right)\Bigg] +\mathcal{O}\left(x\right)\,, \\
H^{22}_{\rm{SO}} &= x^{3/2}\Bigg\{-\frac{4}{3}  \delta\,  \left(\bm{\hat{\ell}} \cdot \bm{\chi}^{c}_{a}\right)-\frac{4}{3} (1-\nu ) \left(\bm{\hat{\ell}} \cdot \bm{\chi}^{c}_{s}\right)+e
   \Bigg[\left(-\frac{7}{3}  \mathrm{e}^{-i l}-\frac{8}{3} \mathrm{e}^{i l}\right)
   \delta \,  \left(\bm{\hat{\ell}} \cdot \bm{\chi}^{c}_{a}\right)+\bigg[-\frac{7}{3}  \mathrm{e}^{-i l} \left(1-\frac{5 \nu
   }{14}\right) \, \nonumber \\
   &-\frac{8}{3} \mathrm{e}^{i l} \left(1-\frac{13 \nu
   }{16}\right)\bigg] \left(\bm{\hat{\ell}} \cdot \bm{\chi}^{c}_{s}\right)\Bigg]\Bigg\}+\mathcal{O}\left(x^{5/2}\right)\,, \\
H^{22}_{\rm{SS}} &=\frac{x^2}{8}\Bigg\{ \bigg[8+15\, e \left(\mathrm{e}^{-i l}+\mathrm{e}^{i l}\right)\bigg] 
   \bigg[2 \left(\bm{\hat{\ell}} \cdot \bm{\chi}^{c}_{a}\right) \, \left(\bm{\hat{\ell}} \cdot \bm{\chi}^{c}_{s}\right)\, \Big[(1-2 \nu ) \kappa _a+\delta \, \kappa
   _s\Big]+\left(\bm{\hat{\ell}} \cdot \bm{\chi}^{c}_{a}\right){}^2 \Big[-2 \nu +\delta\,  \kappa
   _a+(1-2 \nu) \, \nonumber \\ 
   & \times \kappa _s\Big]+\left(\bm{\hat{\ell}} \cdot \bm{\chi}^{c}_{s}\right){}^2
   \Big[2 \nu +\delta \, \kappa _a +(1-2 \nu)  \kappa
   _s\Big]\bigg]\Bigg\} +\mathcal{O}\left(x^{5/2}\right)\,, \\
H^{21}_{{\rm{NS}}} &= -\frac{\delta}{3}  \, \sqrt{x} \Bigg[1+e\,\left(\mathrm{e}^{i l} + \mathrm{e}^{-i l}\right)\Bigg] +\mathcal{O}\left(x^{3/2}\right)\,, \\
H^{21}_{\rm{SO,1PN}} &= \frac{x}{2}\, \Bigg\{\left(\bm{\hat{\ell}} \cdot \bm{\chi}^{c}_{a}\right) +\delta\, \left(\bm{\hat{\ell}} \cdot \bm{\chi}^{c}_{s}\right) +e  \bigg[\left[\left(\bm{\hat{\ell}} \cdot \bm{\chi}^{c}_{a}\right) +\delta\, \left(\bm{\hat{\ell}} \cdot \bm{\chi}^{c}_{s}\right)\right]\,\left(\mathrm{e}^{i l} + \mathrm{e}^{-i l}\right)\bigg]\Bigg\} +\mathcal{O}\left(x^2\right)\,, \\
H^{21}_{\rm{SO,2PN}} &= x^2 \Bigg\{\frac{1}{6} \left(1-\frac{205 \nu }{7}\right)\left(\bm{\hat{\ell}} \cdot \bm{\chi}^{c}_{a}\right) +\frac{\delta}{6}
    \left(1-\frac{33 \nu }{7}\right) \left(\bm{\hat{\ell}} \cdot \bm{\chi}^{c}_{s}\right) +e \Bigg[\bigg[\frac{5 }{6} \mathrm{e}^{-i l}\left(1-\frac{165\nu}{14}\right)+\frac{13 }{3} \mathrm{e}^{i l} \left(1-\frac{991\nu}{364}\right)\bigg] \nonumber \\
   & \times \left(\bm{\hat{\ell}} \cdot \bm{\chi}^{c}_{a}\right) + \delta 
   \left[\frac{5 }{6} \mathrm{e}^{-i l} \left(1-\frac{169\nu}{70}\right)+\frac{13 }{3} \mathrm{e}^{i l}\left(1-\frac{11\nu}{28}\right) \right] \left(\bm{\hat{\ell}} \cdot \bm{\chi}^{c}_{s}\right)\Bigg]\Bigg\}+\mathcal{O}\left(x^{5/2}\right)\,, \\
\label{eq:20NS_ellp_leading_e}
 H^{20}_{\rm{NS}} &=  -\frac{1}{2\sqrt{6}} e \left(\mathrm{e}^{i l}+\mathrm{e}^{-i l}\right)+\mathcal{O}\left(x\right)\,, \\
 H^{20}_{\rm{SO}} &=  -\frac{1}{3\sqrt{6}}x^{3/2}\bigg[e\left(\mathrm{e}^{i l} + \mathrm{e}^{ -i l}\right)\bigg] \bigg[7 \delta \left(\bm{\hat{\ell}} \cdot \bm{\chi}^{c}_{a}\right) +7\left(\bm{\hat{\ell}} \cdot \bm{\chi}^{c}_{s}\right)\left(1-\frac{5\nu}{7}\right) \bigg]+\mathcal{O}\left(x^{5/2}\right)\,, \\ 
\label{eq:20SS_ellp_leading_e}
 H^{20}_{\rm{SS}} &= \frac{1}{4} \sqrt{\frac{3}{2}} x^2 e  \left(\mathrm{e}^{ i l}+\mathrm{e}^{-i l}\right)  \Bigg\{2 \left(\bm{\hat{\ell}} \cdot \bm{\chi}^{c}_{a}\right)\left(\bm{\hat{\ell}} \cdot \bm{\chi}^{c}_{s}\right) \bigg[\delta  \kappa_s + \kappa _a\left(1-2\nu\right)\bigg]+\left(\bm{\hat{\ell}} \cdot \bm{\chi}^{c}_{a}\right)^{2} \bigg[-2\nu +\delta \kappa_a  +\kappa _s\left(1-2\nu\right)\bigg]\, \nonumber \\
   &  +\left(\bm{\hat{\ell}} \cdot \bm{\chi}^{c}_{s}\right)^{2} \bigg[2\nu+\delta \kappa _a 
  + \kappa _s\left(1-2\nu\right) \bigg]\Bigg\}+\mathcal{O}\left(x^{5/2}\right)\,, \\
H^{33}_{{\rm{NS}}} &=- \frac{3\sqrt{15}}{4 \sqrt{14}} \delta\, \sqrt{x}\, \Bigg[1+e \left(\frac{5\, \mathrm{e}^{-i l}}{9}+\frac{5\, \mathrm{e}^{i l}}{3}\right)\Bigg] +\mathcal{O}\left(x^{3/2}\right) \,, \\
H^{33}_{\rm{SO}} &= x^{2}\Bigg\{\frac{3}{2} \sqrt{\frac{15}{14}} \left(1-\frac{19 \nu }{4}\right) \left(\bm{\hat{\ell}} \cdot \bm{\chi}^{c}_{a}\right)+\frac{3}{2} \sqrt{\frac{15}{14}} \delta  \left(1-\frac{5 \nu
   }{4}\right) \left(\bm{\hat{\ell}} \cdot \bm{\chi}^{c}_{s}\right)+e \Bigg[\bigg[\frac{15}{4}
   \sqrt{\frac{15}{14}} \mathrm{e}^{i l} \left(1-\frac{139 \nu
   }{30}\right)+\frac{5}{4} \sqrt{\frac{35}{6}} \mathrm{e}^{-i l}\, \nonumber \\
   & \times \bigg(1-\frac{327 \nu }{70}\bigg)\bigg]\left(\bm{\hat{\ell}} \cdot \bm{\chi}^{c}_{a}\right)+\delta 
   \bigg[\frac{5}{4} \sqrt{\frac{35}{6}} \mathrm{e}^{-i l} \left(1-\frac{9 \nu
   }{14}\right)+\frac{15}{4} \sqrt{\frac{15}{14}} \mathrm{e}^{i l}
   \left(1-\frac{11 \nu }{10}\right)\bigg] \left(\bm{\hat{\ell}} \cdot \bm{\chi}^{c}_{s}\right)\Bigg]\Bigg\} +\mathcal{O}\left(x^{5/2}\right)\,, \\
H^{32}_{{\rm{NS}}} &= \frac{\sqrt{5}}{3 \sqrt{7}} \left(1-3 \nu \right)\, x  \Bigg[1 +\frac{1}{8} e  \left(11\, \mathrm{e}^{-i l}+13\, \mathrm{e}^{ i l}\right)\Bigg] +\mathcal{O}\left(x^2\right)\,, \\
H^{32}_{\rm{SO}} &= \frac{\sqrt{5}}{6\sqrt{7}}\,\nu \,x^{3/2} \Bigg[8 +e \left(11\, \mathrm{e}^{-i l} +13\, \mathrm{e}^{i l} \right)\Bigg]\,\left(\bm{\hat{\ell}} \cdot \bm{\chi}^{c}_{s}\right) +\mathcal{O}\left(x^{5/2}\right)\,, \\
H^{31}_{{\rm{NS}}} &= -\frac{1}{12 \sqrt{14}}\,\delta\, \sqrt{x} \Bigg[1-e \left(5\,\mathrm{e}^{-i l}-\mathrm{e}^{i l}\right)\Bigg] +\mathcal{O}\left(x^{3/2}\right)\,, \\
H^{31}_{\rm{SO}} &= x^{2}\Bigg\{\frac{1}{6
   \sqrt{14}}\left(1-\frac{11 \nu }{4}\right) \left(\bm{\hat{\ell}} \cdot \bm{\chi}^{c}_{a}\right)+\frac{\delta}{6 \sqrt{14}}  \left(1-\frac{13 \nu }{4}\right) \left(\bm{\hat{\ell}} \cdot \bm{\chi}^{c}_{s}\right)+e \Bigg[\bigg[\frac{1}{12} \sqrt{\frac{7}{2}} \mathrm{e}^{i
   l} \left(1-\frac{51 \nu }{14}\right)+\frac{25 }{12 \sqrt{14}} \mathrm{e}^{-i l}  \, \nonumber \\
   & \times \left(1-\frac{181 \nu }{50}\right)\bigg]\left(\bm{\hat{\ell}} \cdot \bm{\chi}^{c}_{a}\right) +\delta  \bigg[\frac{1}{12} \sqrt{\frac{7}{2}} \mathrm{e}^{i l}
   \left(1-\frac{17 \nu }{14}\right)+\frac{25 }{12 \sqrt{14}}\mathrm{e}^{-i l} \left(1-\frac{63
   \nu }{50}\right)\bigg] \left(\bm{\hat{\ell}} \cdot \bm{\chi}^{c}_{s}\right)\Bigg]\Bigg\} +\mathcal{O}\left(x^{5/2}\right)\,, \\
 H^{30}_{\rm{NS}} &= \frac{1}{4 \sqrt{42}}x\,e \left(\mathrm{e}^{i l}-\mathrm{e}^{-i l}\right) \left(1-3 \nu \right)  +\mathcal{O}\left(x^2\right)\,, \\
H^{30}_{\rm{SO}} &= \frac{1}{\sqrt{42}}\nu x^{3/2} e  \left(\mathrm{e}^{ i l}-\mathrm{e}^{-i l}\right) \left(\bm{\hat{\ell}} \cdot \bm{\chi}^{c}_{s}\right) +\mathcal{O}\left(x^{5/2}\right)\,, \\
H^{43}_{{\rm{NS}}} &=- \frac{9 }{4 \sqrt{70}} \,\delta \left(1-2 \nu \right) \, x^{3/2} \Bigg[1+\frac{1}{27} e  \left(47\mathrm{e}^{-i l}+57 \mathrm{e}^{ i l}\right)\Bigg] +\mathcal{O}\left(x^{5/2}\right)\,, \\
H^{43}_{\rm{SO}} &= \frac{\sqrt{5}}{8 \sqrt{14}}\,\nu\, x^{2} \, \Bigg\{9 \bigg[\left(\bm{\hat{\ell}} \cdot \bm{\chi}^{c}_{a}\right) -\delta\,\left(\bm{\hat{\ell}} \cdot \bm{\chi}^{c}_{s}\right)\bigg]  +e \left(19\,\mathrm{e}^{i l} +\frac{47}{3}\mathrm{e}^{-i l} \right) \bigg[\left(\bm{\hat{\ell}} \cdot \bm{\chi}^{c}_{a}\right) -\delta\,\left(\bm{\hat{\ell}} \cdot \bm{\chi}^{c}_{s}\right)\bigg]\Bigg\} +\mathcal{O}\left(x^{5/2}\right)\,, \\
H^{41}_{{\rm{NS}}} &=- \frac{1}{84 \sqrt{10}} \delta \left(1-2 \nu \right) x^{3/2}\Bigg[1-e  \left(9\,\mathrm{e}^{-i l}-\mathrm{e}^{ i l}\right)\Bigg] +\mathcal{O}\left(x^{5/2}\right)\,, \\
H^{41}_{\rm{SO}} &=\frac{\sqrt{5}  }{168 \sqrt{2}} \nu \, x^2 \Bigg\{\bigg[\left(\bm{\hat{\ell}} \cdot \bm{\chi}^{c}_{a}\right) -\delta\,\left(\bm{\hat{\ell}} \cdot \bm{\chi}^{c}_{s}\right)\bigg]  -e \left(9 \,\mathrm{e}^{-i l}-\mathrm{e}^{i l}\right) \bigg[\left(\bm{\hat{\ell}} \cdot \bm{\chi}^{c}_{a}\right) -\delta\,\left(\bm{\hat{\ell}} \cdot \bm{\chi}^{c}_{s}\right)\bigg]\Bigg\} +\mathcal{O}\left(x^{5/2}\right)\,.
\end{align}
\end{subequations}
\end{widetext}
Note that the PN parameter ($x$) used above is related to the PN parameter ($y$) as $x=y^2 (1-e^2)$. Note that we continue to use the parameter $y$ for the arbitrary eccentricity results following the suggestion of \cite{Klein:2018ybm} as can be seen in Appendix~\ref{sec:qk_arb}. When writing eccentricity expanded results, however, we change to the parameter $x$ which is defined as 
\begin{align}
x&=\Bigg[\frac{G\,M\,(1+k)\,n}{c^3}\Bigg]^{2/3}\,,
\label{eq:x2n}
\end{align} 
where different symbols have been defined above. This choice is to provide results in a notation consistent with the 3PN nonspinning counterparts listed in \cite{Boetzel:2019nfw, Ebersold:2019kdc} without running into problems related to its use in arbitrary eccentricity results as we treat eccentricity as a small parameter. Note also, while eccentric anomaly ($u$) has been used while writing arbitrary eccentricity results, we use mean anomaly ($l$) again to be consistent with the choices of \cite{Boetzel:2019nfw, Ebersold:2019kdc}. Note that this requires one to make use of $u$-$l$ relation obtained by solving the Kepler equation that connects the two.  A General solution to the Kepler equation for nonspinning systems has been outlined in \cite{Boetzel:2019nfw} and can be used to write the desired relation to arbitrary powers of eccentricity. For our current purposes (SO and SS contributions through 2PN), we only require 1PN relation with leading\footnote{For the results in the Supplemental Material~\cite{kp:suppl} the $u$-$l$ relation to $\mathcal{O}(e^7)$ is used.} eccentricity correction. It reads
\begin{align}
      u &= \Big[l + e\, \sin l +\mathcal{O}\left(e^2\right) \Big]+ \mathcal{O}\left(x^2\right)\,.
      \label{eq:u-l}
\end{align}
\vskip 5pt
\noindent
\section{quasicircular limit}
\label{sec:circlim}
Both QK and general orbit expressions for various modes reduce to circular versions available in literature \cite{Arun:2008kb, Buonanno:2012rv}. While the circular limit of elliptical orbit results is listed in Sec.~\ref{sec:qk} and Appendix~\ref{sec:qk_arb} by setting $e\rightarrow0$, general orbit results can be reduced to the circular case by setting $\dot{r}$=0 and $\dot{\phi}=\omega_\phi=n$.\footnote{Leading corrections to $\dot{r}$ appear only at 2.5PN order.} For the circular case, Eq.~\eqref{eq:x2n} becomes
\begin{align}
x&=\left(\frac{G M n}{c^3}\right)^{2/3}\,.
\label{eq:x2n_circ}
\end{align}
The dependence on the radial separation $r$ can then be expressed in terms of $n$ [and hence in terms of $x$ using Eq.~\eqref{eq:x2n_circ}].  
For binaries in circular orbits, the radial separation $r$ in terms of $x$ with spin corrections to 2PN is given in Refs.~\cite{Blanchet:2006gy,Bohe:2015ana}. We only reproduce NS, SO, and SS pieces that are needed to reproduce the limits. They read

\begin{align}
\gamma=x\Big[\gamma^{\rm NS}+\gamma^{\rm SO}+\gamma^{\rm SS}\Big]\,,
\label{eq:gamma}
\end{align}
\vskip 2 pt
\noindent
where individual pieces read
\begin{subequations}
    \begin{align}
        \gamma^{\rm NS}&=1+x\,\left(1-\frac{\nu}{3} \right)+\mathcal{O}\left(x^2\right)\,,\\ 
        \gamma^{\rm SO}&=\frac{1}{3}\,x^{3/2} \Bigg[5\, \left(\bm{\hat{\ell}}\cdot\bm{s}_{c}\right)+3 \,\delta \, \left(\bm{\hat{\ell}}\cdot\bm{\sigma}_{c}\right)\Bigg]+\mathcal{O}\left(x^{5/2}\right)\,,\\ 
        \gamma^{\rm SS}&= - \frac{1}{2}x^2 \left(\bm{s}^{+}_{0}\cdot\bm{s}^{-}_{0}\right)+\mathcal{O}\left(x^3\right)\,.
    \end{align}
    \label{eq:gamma_pieces}
\end{subequations}
Here, $\gamma$ is a another PN parameter and is related to the binary's radial separation ($r$) as
\begin{align}
    \gamma=\left(\frac{GM}{c^2r}\right)\,.
    \label{eq:gamma_to_r}
\end{align}
\vskip 5 pt
With $\dot{r}=0$ and both ($r$, $\dot{\phi}$) expressed in terms of $x$ through Eqs.~\eqref{eq:x2n_circ}-\eqref{eq:gamma_pieces},  the circular limit of general orbit results can be written completely in terms of the PN parameter $x$. We have verified that our general orbit results presented in Sec.~\ref{sec:genOrb_hlm} reduce to the circular versions presented in \cite{Arun:2008kb,Buonanno:2012rv}. It may be worth noting that the $(2, 0)$ mode at the leading (nonspinning) order also has nonoscillatory memory contributions originating from memory integrals appearing in the expression for the mass quadrupole moment \cite{Ebersold:2019kdc}. Since we do not discuss hereditary contributions in this work, we ignore these terms here. A Null circular limit for the expression of the $(2,0)$ mode [see Eq.~\eqref{eq:20NS_ellp_leading_e} and Eq.~\eqref{eq:20NS_arb_leading_e}] is indicative of the same. Note also, the circular orbit limit in terms of (anti)symmetric combinations of dimensionless spin parameters ($\bm{\chi}^{c}_s, \bm{\chi}^{c}_a, \kappa_s, \kappa_a$) is automatically extracted from Eq.~\eqref{eq:Hlm_qK_leading_e}. 
\section{Summary and Discussion}
\label{sec:conclusion}
In this paper, we presented computations of spin-orbit and spin-spin effects in spherical harmonic modes of gravitational waveforms through 2PN order within the generalized framework of \cite{Kidder:2007rt, Buonanno:2012rv}. Section \ref{sec:PNinputs} lists the inputs required for these computations.  The first set of results include SO and SS contributions to each mode contributing to the waveform through 2PN order for compact binaries in \textit{general} orbits with\textit{ nonprecessing} spins and have been listed in Sec.~\ref{sec:genOrb_hlm} in terms of radial ($r, \dot{r}$) and angular ($\phi, \dot{\phi}$) variables for position and velocity as well as in terms of suitable combinations of individual spins. We choose to express these results in terms of two sets of spin variables. Equation~\eqref{eq:genOrb_Hlm} presents results in terms of spin variables defined in \cite{Buonanno:2012rv} so as to reproduce exact expressions of \cite{Arun:2008kb, Buonanno:2012rv} when the circular limit is taken.\footnote{In fact,  we use a dimensionless version of these variables defined in Eq.~\eqref{eq:spinVarRedefine}.} Additionally, following \cite{Mishra:2016whh} we also list our results as Eq.~(\ref{eq:genOrb_Hlm_chia_chis}) in terms of symmetric ($\bm{\chi}^{c}_{s}$) and antisymmetric ($\bm{\chi}^{c}_{a}$) combinations of dimensionless spin variables $\bm{\chi}^{c}_{1, 2}={\bm{S^{c}}_{1,2}}/G\,m^{2}_{1,2}$. In this presentation of results, the dependence on spin-induced parameters for individual binary components ($\kappa_{1,2}$) appearing in the SS part for the dominant ($\ell$=$2$, $m$=$2$) mode also becomes explicit.\footnote{Note that this is also true for the {($\ell$=$2$, $m$=0)} mode.} Again, we choose to write the final expressions in terms of a symmetric and antisymmetric combination of ($\kappa_{1,2}$) defined above. Note that through the 2PN order not all modes have spin corrections and only ($\ell$, $|m|$)=($(2,2), (2,1), (2,0), (3,3), (3,2), (3,1), (3,0), (4,3),\,\\ (4,1)$) modes contribute. We do not list the modes that do not have spin corrections through 2PN when presenting the results in this paper. Note also that through 2PN only the ($\ell$=$2$, $|m|$=$2$) mode has SS contributions and hence a dependence on spin-induced parameters.\footnote{ Note again that the leading $m=0$ mode $(\ell=2, m=0)$ also has SS contribution as can be seen in Eq.~\eqref{eq:20SSGen}, \eqref{eq:20SSGen_chi}, \eqref{eq:20SS_ellp_leading_e}.} Additionally, up to 2PN order only the (2, 1) mode has leading (at 1PN) and subleading (at 2PN) SO contribution, for other modes sub-leading terms appear beyond 2PN. Finally, the circular orbit limit of our general orbit results can be obtained using the transformations discussed in Sec.~\ref{sec:circlim}. We have verified that our results reduce to circular results of \cite{Arun:2008kb, Buonanno:2012rv} up to 2PN.
 \vskip 5pt
Next, we specialize to the case of elliptical orbits using the QK representation of \cite{Klein:2010ti, Klein:2018ybm} for spinning compact binaries. These results are listed in Sec.~\ref{sec:qk} and in Appendix \ref{sec:qk_arb}. Here, we have used the relations connecting \textit{general} orbit variables to those characterizing \textit{elliptical} orbits (listed in Appendix~\ref{sec:qk_orbdyn}) in the expressions for each mode written in terms of $r, \,\dot{r}$ and $\dot{\phi}$ and given by Eq.~\eqref{eq:genOrb_Hlm}. While the results presented in Sec.~\ref{sec:genOrb_hlm} are useful for orbits with small eccentricities, arbitrary eccentricity results have been pushed to Appendix~\ref{sec:qk_arb} to avoid any disruption of the flow of the paper as these expressions run over a  couple of pages. Similar to the general orbit results, we only list modes which have spin contributions through 2PN with SO, SS, and leading NS pieces presented separately. Also, we choose to present arbitrary eccentricity results in terms of spin combinations ($\beta, \gamma_1$) defined in \cite{Klein:2018ybm}, although, in our presentation we scale them with appropriate factors of $G$ and $M$ (omitted in \cite{Klein:2018ybm}) to make them dimensionless\footnote{We redefine $(\beta, \gamma_1)$ of \cite{Klein:2018ybm} as $\frac{{\beta(a,b)}}{G\,M}\rightarrow\beta(a,b)$ and $\frac{\gamma_{1}}{G^{2}\,M^{2}\,}\rightarrow\gamma_{1}$.} and reexpressed as Eq.~\eqref{eq:exp_beta_gamma1} in terms of dimensionless spin variables ($\bm{\chi}_s,\bm{\chi}_a$) for compactness. Further, these results have been expressed as explicit functions of orbital time eccentricity ($e$), eccentric anomaly ($u$), and the PN parameter $(y)$ introduced in \cite{Klein:2018ybm}. The small eccentricity results presented in Sec.~\ref{sec:qk} include leading-order eccentric corrections to circular results, expressed explicitly in terms of another PN parameter ($x$) related to $y$ and $e$ (see a discussion in Sec.~\ref{sec:qk}). Other parameters that appear include orbital time eccentricity $e$, mean anomaly ($l$), and spin parameters $\bm{\chi}^{c}_s$ and $\bm{\chi}^{c}_a$. Higher order corrections to eccentricity [to $\mathcal{O}(e^7)$] ofthese results together with nonspinning contributions to 3PN are included in the Supplemental Material~\cite{kp:suppl}.
 \vskip 5pt
This paper lists explicitly the spin contributions to different spherical harmonic modes through 2PN order for binaries in general orbits. These together with the nonspinning results of \cite{Mishra:2015bqa} provide state-of-the-art results for compact binaries in orbits of arbitrary shape and nature. Moreover, elliptical orbit results presented in Appendix~\ref{sec:qk_arb} together with the related results of \cite{Mishra:2015bqa} are again the most up-to-date versions of PN eccentric waveforms valid for arbitrary eccentricities. We recall here that the nonspinning results of \cite{Mishra:2015bqa} do not include the contributions from hereditary terms and were computed later in follow-up papers \cite{Boetzel:2017zza, Ebersold:2019kdc} in the small eccentricity approximation. In fact, as was argued in \cite{Boetzel:2017zza}, closed-form expressions for hereditary terms (for any noncircular orbit) can only be obtained under the small eccentricity approximation. While a hereditary contribution associated with spins appears in waveforms beyond the 2PN order and is not our immediate concern, future works involving these contributions shall also involve an expansion of eccentricity parameter. (See however a very recent work \cite{Henry:2022dzx} where these contributions were computed for binaries in circular orbits.) References~\cite{Boetzel:2017zza, Ebersold:2019kdc} provide complete 3PN contributions to GW modes for eccentric binaries with nonspinning components to $\mathcal{O}(e^7)$. Our current computations add to these results the spin contributions through 2PN order for  the nonprecessing case again to $\mathcal{O}(e^7)$. Complete results with a 3PN accurate nonspinning part and a 2PN spinning part valid for nonprecessing systems are being provided as Supplemental Material~\cite{kp:suppl} and should be useful for systems with small to moderate eccentricities.     
 \vskip 5pt
 Results presented here together with the nonspinning results presented in earlier works \cite{Mishra:2015bqa, Boetzel:2017zza, Ebersold:2019kdc} should be useful for accurate comparisons with state-of-the art eccentric numerical relativity simulations  for spinning BBHs \cite{Joshi:2022ocr} and should set the platform for construction of complete inspiral-merger-ringdown models for eccentric, nonprecessing BBH models including the effect of higher order modes. Note, however, that in this work we only discuss the computation of spin effects in the amplitude of each relevant mode. The spinning corrections to the secular phase can be computed using the energy and flux expressions given in Ref.~\citep{Klein:2018ybm} following an approach similar to \cite{Moore:2016qxz} which explicitly computes the GW secular phase to 3PN order assuming eccentricity as a small parameter~\cite{Paul-etal-2022}.
\vskip 5pt
Finally, these spherical harmonic mode expressions can be combined using the usual basis of spherical harmonics to obtain polarization waveforms that are typically used in data-analysis applications. The combined effect of spins, eccentricity, and higher modes computed here should help reduce systematic biases in various parameter estimation analyses for inspiral dominated signals. Moreover, these also could be used to obtain an analytical frequency domain model (say based on stationary phase approximation) that can be used for performing detailed parameter estimation with eccentricity as a parameter.
\section*{ACKNOWLEDGEMENTS}
We thank Guillaume Faye and K. G. Arun for useful discussions. We thank the referee and Mohammed Khalil for pointing out a possible error in our computations of the $(2,1)$ mode. We sincerely thank Quentin Henry, Guillaume Faye, and Mohammed Khalil for the discussions that helped us resolve the discrepancy. We also thank members of the Gravitation \& Cosmology group at IIT Madras for numerous insightful discussions. All computations were performed using \textit{Mathematica} and the tensor algebra package, xA{\tiny{CT}}~\citep{XAct:xTensor}. 
\appendix
\section{ORBITAL DYNAMICS}
\label{sec:qk_orbdyn}
General orbit variables $(r, \dot{r}, \phi, \dot{\phi})$ are written explicitly in terms of parameters characterizing the binary's orbital dynamics in elliptical orbits $(y, e, u)$. Only relevant pieces (desired for investigations performed here) are computed starting with the QK representation of Ref.~\cite{Klein:2018ybm}. These read as
\begin{subequations}
\label{eq:exp_r_rdot_phi_phidot_Struc}
 \begin{align}
    r &=  r^{{\rm{NS}}} + r^{{\rm{SO}}} + r^{{\rm{SS}}}\,, \\
    \dot{r} &= \dot{r}^{{\rm{NS}}} + \dot{r}^{{\rm{SO}}} + \dot{r}^{{\rm{SS}}}\,, \\
    \phi &= \phi^{{\rm{NS}}} + \phi^{{\rm{SO}}} + \phi^{{\rm{SS}}}\,, \\
    \dot{\phi} &=  \dot{\phi}^{{\rm{NS}}} + \dot{\phi}^{{\rm{SO}}} + \dot{\phi}^{{\rm{SS}}}\,,
\end{align}   
\end{subequations}
where
\begin{subequations}
\label{eq:exp_r_NS_SO_SS}
\begin{align}
 r_{{\rm{Newt}}} &= \frac{G M (1-e \cos u)}{c^2 \left(1-e^2\right) y^2}\,, \\ 
 r^{{\rm{NS}}} &= r_{{\rm{Newt}}}\Bigg\{1+y^{2}\bigg[\frac{1}{3} \left[-3+\nu +9 e^2 \left(1-\frac{\nu
   }{9}\right)\right] \, \nonumber \\
   &-\frac{4 e \left(1-e^2\right) \cos u
  }{1-e \cos u}\, \left(1-\frac{3 \nu }{8}\right)\bigg]\Bigg\}\,, \\
 r^{{\rm{SO}}} &= r_{{\rm{Newt}}}\, y^3 \Bigg[-\frac{e\,\left(1- e^{2}\right) \cos u}{2 \left(1-e \cos u\right)} \beta (4,2) \, \nonumber \\
 &+\beta \left(\frac{2}{3}+2
   e^2,1+e^2\right)\Bigg]\,, \\
 r^{{\rm{SS}}} &= r_{{\rm{Newt}}}\, y^4 \Bigg\{\gamma_{1} \bigg[\frac{1+e^2}{2}-\frac{e\,\left(1- e^{2}\right) \cos u}{2 \left(1-e \cos u\right)}\bigg]\Bigg\}\,,
 \end{align}
 \end{subequations}
 \begin{subequations}
\label{eq:exp_rdot_NS_SO_SS}
 \begin{align}  
 \dot{r}_{{\rm{Newt}}} &= \frac{c\, e\, \sqrt{1-e^2}\, \sin u\, y}{(1-e \cos u)}\,, \\
  \dot{r}^{{\rm{NS}}} &= \dot{r}_{{\rm{Newt}}}\Bigg\{1+y^2 \left[-e^2+\left(-\frac{7  }{6}+\frac{7 e^2  }{6}\right)\nu\right]\Bigg\}\,, \\
 \dot{r}^{{\rm{SO}}} &= \dot{r}_{{\rm{Newt}}}\,y^3 \Bigg[\left(\frac{1}{2}-\frac{e^2}{2}\right) \beta \left(4,2\right)-\beta
   \left(4,3\right) \, \nonumber \\
   &+\beta \left(\frac{2}{3}+2 e^2,1+e^2\right)\Bigg]\,, \\
  \dot{r}^{{\rm{SS}}} &= -\dot{r}_{{\rm{Newt}}}\,\frac{\gamma_{1}}{2} y^4\,,
  \end{align}
 \end{subequations}
 \begin{subequations}
 \label{eq:exp_phi_NS_SO_SS}
 \begin{align}
  \phi^{{\rm{NS}}} &=  \Bigg\{V\left(e,u\right) + y^2 \bigg[3 V\left(e,u\right)+\frac{4e \sqrt{1-e^2}  \sin u}{
    \left(1-e \cos u\right)}\, \nonumber \\
    &\left(1-\frac{\nu}{4} \right)\bigg]\Bigg\}\,, \\
  \phi^{{\rm{SO}}} &= y^3 \Bigg[V\left(e,u\right) \beta \left(4,3\right)+\frac{e \sqrt{1-e^2} \sin u }{2 \left(1- e \cos u\right)}\beta
   \left(4,4\right)\Bigg]\,, \\
   \phi^{{\rm{SS}}} &= y^4 \Bigg\{\gamma_{1} \bigg[\frac{3}{2} V\left(e,u\right) + \frac{e \sqrt{1-e^2} \sin u}{\left(1-e
   \cos u \right)} \bigg]\Bigg\}\,,
   \end{align}
 \end{subequations}
with
\begin{align}
  V(e,u) &= 2\, \mathrm{arctan}\left[\sqrt{\frac{\left(1+e\right)}{\left(1-e\right)}} \mathrm{tan}\frac{u}{2}\right]\,,
  \label{eq:exp_V}
\end{align} 
 \begin{subequations}
 \label{eq:exp_phidot_NS_SO_SS}
 \begin{align}
 \dot{\phi}_{{\rm{Newt}}} &= \frac{c^3 \left(1-e^2\right)^2 y^3}{G M \left(1-e \cos u\right)^2}\,, \\
  \dot{\phi}^{{\rm{NS}}} &= \dot{\phi}_{{\rm{Newt}}}\Bigg\{1+ y^2\,\bigg[\frac{ 4 e  \left(\cos
   u - e\right)}{\left(1-e \cos u\right)}\left(1-\frac{\nu }{4}\right)\bigg]\Bigg\}\,, \\
  \dot{\phi}^{{\rm{SO}}} &= \dot{\phi}_{{\rm{Newt}}}\,y^3 \Bigg\{\bigg[-\frac{1}{2}+\frac{1- e^2}{2 \left(1-e \cos u\right)}\bigg] \beta \left(4,4\right)\Bigg\}\,, \\
  \dot{\phi}^{{\rm{SS}}} &= \dot{\phi}_{{\rm{Newt}}}\,y^4 \Bigg\{\gamma_{1} \bigg[-1+\frac{1-e^2}{\left(1-e \cos
   u\right)}\bigg]\Bigg\}\,.
\end{align}
\end{subequations}
In the above, 
\begin{subequations}
\label{eq:exp_beta_gamma1}
\begin{align}
    \beta\left(a,b\right) &= -\, \Bigg\{ a \Big[\left(1-2\nu\right)\,\left(\bm{\hat{\ell}} \cdot \bm{\chi}_{s}\right)+\delta\,\left(\bm{\hat{\ell}} \cdot \bm{\chi}_{a}\right)   \Big] \, \nonumber \\
    &+2 b \nu\,\left(\bm{\hat{\ell}} \cdot \bm{\chi}_{s}\right)\Bigg\}, \\
    \gamma_{1} &= \Bigg\{\Big[\delta\, \kappa_{a} + \kappa_{s} (1-2 \nu )-2 \nu \Big]\left(\bm{\hat{\ell}} \cdot \bm{\chi}_{a}\right)^{2} +2\Big[ \delta\, \kappa_{s} \nonumber \\
    &+\kappa_{a} \left(1-2 \nu \right)\Big]\left(\bm{\hat{\ell}} \cdot \bm{\chi}_{a}\right) \left(\bm{\hat{\ell}} \cdot \bm{\chi}_{s}\right)+\Big[\delta\, \kappa_{a} + \kappa_{s}\, \nonumber \\
    & \times \left(1-2 \nu \right) +2 \nu \Big] \left(\bm{\hat{\ell}} \cdot \bm{\chi}_{s}\right)^{2} \Bigg\}.
\end{align}
\end{subequations}
\vspace{0.01cm}
\section{SPHERICAL HARMONIC MODES WITH SPIN EFFECTS TO 2PN: ARBITRARY ECCENTRICITY CASE}
\label{sec:qk_arb}
Spherical harmonic modes with spin effects to 2PN, valid for elliptical orbits of arbitrary eccentricity, in terms of parameters $(e, u, y)$ read
\begin{equation}
\label{eq:qkhlm}
    h^{\ell m} = -i^{2m} \frac{8\,G\,M\,\nu\,\left(1-e^2\right)\,y^2}{c^{2}\,R}\,\sqrt{\frac{\pi}{5}}\,\mathrm{e}^{-im\phi}\,H^{\ell m}
\end{equation}
where
\begin{widetext}
\begin{subequations}
\label{eq:qkHlm}
\begin{align}
 H^{22}_{\rm{NS}} &= \frac{1}{4
   \left(1-e \cos u\right)^2}\Bigg(4-3 e^2-2 e \cos u+e^2 \cos 2 u+4 i\, e \sqrt{1-e^2} \sin u\Bigg) +\mathcal{O}\left(y^2\right) \,, \\
 H^{22}_{{\rm{SO}}} &= -\frac{y^3}{6 \left(1-e \cos u\right)^3}\Bigg\{ 2 \delta \, \left(\bm{\hat{\ell}} \cdot \bm{\chi}^{c}_{a}\right) \Bigg[4-11 e^2+3 e^4+e \left(3+4 e^2\right)
   \cos u-4 e^2 \cos 2 u+e^3 \cos 3 u+ \sqrt{1-e^2}\bigg[ \Big(i\, e  \, \nonumber \\
   &-3
   i\, e^3 \Big) \sin u+ i\, e^2  \sin 2 u\bigg]\Bigg]+\left(\bm{\hat{\ell}} \cdot \bm{\chi}^{c}_{s}\right) \Bigg[8-22 e^2+6 e^4-8 \nu +10 e^2 \nu +2 e^4 \nu +\Big(2 e^3 -e^3 \nu\Big)  \cos 3 u -8 e^2 \, \nonumber \\
   & \times \cos 2 u \left(1-\frac{\nu
   }{2}\right)+\cos u \left[6 e (1+\nu )+8 e^3 \left(1-\frac{13 \nu
   }{8}\right)\right]+ \sqrt{1-e^2} \bigg[\Big(2 i\, e   -6 i\, e^3 
   -8 i\, e \nu +4 i\, e^3 \nu\Big)  
   \sin u \, \nonumber \\
   &+\Big(2 i\, e^2  +2 i\, e^2 \nu\Big) 
   \sin 2 u\bigg]\Bigg]\Bigg\} +\mathcal{O}\left(y^5\right)\,, \\
   H^{22}_{{\rm{SS}}} &= \frac{y^4}{8 \left(1-e \cos u\right)^3} \Bigg\{\Bigg[8-22 e^2+10 e^4+e \left(6+e^2\right) \cos u-4 e^2 \cos
   2 u+e^3 \cos 3 u-4 i\, e^3 \sqrt{1-e^2} \sin u\, \nonumber \\
   &+2 i\, e^2
   \sqrt{1-e^2} \sin 2 u\Bigg] \Bigg[2 \left(\bm{\hat{\ell}} \cdot \bm{\chi}^{c}_{a}\right) \left(\bm{\hat{\ell}} \cdot \bm{\chi}^{c}_{s}\right) \bigg[\left(1-2
   \nu \right) \kappa _a+\delta \, \kappa _s\bigg]+\left(\bm{\hat{\ell}} \cdot \bm{\chi}^{c}_{a}\right){}^2
   \bigg[-2 \nu +\delta \, \kappa _a+ \left(1-2 \nu\right)  \kappa
   _s\bigg] \, \nonumber \\
   &+\left(\bm{\hat{\ell}} \cdot \bm{\chi}^{c}_{s}\right){}^2 \bigg[2 \nu +\delta \, \kappa
   _a + \left(1-2 \nu\right)  \kappa _s\bigg]\Bigg]\Bigg\} +\mathcal{O}\left(y^5\right)\,, \\
   H^{21}_{\rm{NS}} &= -\frac{\delta\,  \left(1-e^2\right) y}{3 \left(1-e \cos u\right)^2} +\mathcal{O}\left(y^3\right)\,, \\
   H^{21}_{\rm{SO,1PN}} &= \frac{ \left(1-e^2\right) y^2 }{2\left(1-e \cos u\right)^2} \Bigg[\left(\bm{\hat{\ell}} \cdot \bm{\chi}^{c}_{a}\right)+\delta\,\left(\bm{\hat{\ell}} \cdot \bm{\chi}^{c}_{s}\right)\Bigg] +\mathcal{O}\left(y^4\right)\,, \\
   H^{21}_{\rm{SO,2PN}} &= -\frac{ \left(1-e^2\right)y^4}{168 (1-e \cos
   (u))^4} \Bigg\{\delta  \left(\bm{\hat{\ell}} \cdot \bm{\chi}^{c}_{s}\right) \Bigg[-28+1057
   e^2-357 e^4+132 \nu -322 e^2 \nu +190 e^4 \nu +e^2 
   \bigg[203-38 \nu +21 e^2 \, \nonumber \\
   & \times \left(1+\frac{38 \nu }{21}\right)\bigg]\cos 2 u-4 e
   \cos u \bigg[189-24 \nu +35 e^2 \left(1+\frac{24 \nu
   }{35}\right)\bigg] +\sqrt{1-e^2} \bigg[-588 i e+588 i e^3-52 i e \nu \, \nonumber \\
   &+52 i e^3 \nu \bigg] \sin
   u\Bigg]+\left(\bm{\hat{\ell}} \cdot \bm{\chi}^{c}_{a}\right) \Bigg[-28+1057 e^2-357 e^4+820 \nu -2170
   e^2 \nu + 678 e^4 \nu +4 e \bigg[-189
   -35 e^2 \, \nonumber \\
   &\times \left(1-\frac{136 \nu }{35}\right) + 88 \nu\bigg] \cos u +e^2 \bigg[203-270 \nu +21 e^2
   \left(1+\frac{46 \nu }{21}\right)\bigg] \cos 2 u + \sqrt{1-e^2} \bigg[-588 i e+588 i e^3 \, \nonumber \\
   &+ 332 i e \nu - 332 i e^3 \nu \bigg]\sin (u)\Bigg]\Bigg\}+\mathcal{O}\left(y^5\right)\,, \\
   \label{eq:20NS_arb_leading_e}
    H^{20}_{\rm{NS}} &=  -\frac{1}{\sqrt{6}}\frac{e \cos u}{ (1-e \cos u)}+\mathcal{O}\left(y^2\right)\,, \\
    H^{20}_{\rm{SO}} &=  -\frac{e y^3}{3 \sqrt{6} (1-e \cos u)^3} \Bigg\{\delta \left(\bm{\hat{\ell}} \cdot \bm{\chi}^{c}_{a}\right) \bigg[  \left(6 e^3-14 e\right)+2  e^2 \cos 3 u-8   e \cos 2 u+14   \cos u\bigg]+ \left(\bm{\hat{\ell}} \cdot \bm{\chi}^{c}_{s}\right) \, \nonumber \\
   & \times \bigg[6 e^3 (1-\nu )+2 e^2 \left(1-\frac{\nu }{2}\right) \cos 3 u+\left(3 e^2 \nu -10 \nu +14\right) \cos u +10 e \nu -8 e \left(1-\frac{\nu }{2}\right)
   \cos 2 u -14 e\bigg]\Bigg\}+\mathcal{O}\left(y^5\right)\,, \\
    H^{20}_{\rm{SS}} &= \frac{e y^4}{4 \sqrt{6} (1-e \cos u)^3} \Bigg\{\Bigg[\left(e^2+6\right) \cos u+e \left(2 e^2+e \cos 3 u-4 \cos 2 u-6\right)\Bigg] \Bigg[\left(\bm{\hat{\ell}} \cdot \bm{\chi}^{c}_{a}\right)^{2} \bigg[-2\nu+\delta  \kappa_a + \kappa_s \, \nonumber \\
   & \times \left(1-2\nu\right)\bigg] +2 \left(\bm{\hat{\ell}} \cdot \bm{\chi}^{c}_{a}\right)\left(\bm{\hat{\ell}} \cdot \bm{\chi}^{c}_{s}\right) \bigg[\delta  \kappa_s +\kappa_a \left(1-2\nu\right)\bigg]+ \left(\bm{\hat{\ell}} \cdot \bm{\chi}^{c}_{s}\right)^{2} \bigg[2\nu+\delta  \kappa_a + \kappa_s\left(1-2\nu\right)\bigg]\Bigg]\Bigg\}+\mathcal{O}\left(y^5\right)\,, \\
    H^{33}_{\rm{NS}} &= -\frac{ \sqrt{5} \delta\,y }{8 \sqrt{42} \left(1-e
   \cos u\right)^3} \Bigg\{18-28 e^2+10 e^4+\left(1-e^2\right) \bigg[-14 e
   \cos u+ 6 e^2  \cos 2 u\bigg]+\sqrt{1-e^2}\bigg[\Big(20
   i\, e -15 i\, e^3 \Big)  \, \nonumber \\
   &\times \sin u  -4 i\, e^2
    \sin 2 u+ i\, e^3  \sin 3 u\bigg]\Bigg\} +\mathcal{O}\left(y^3\right)\,, \\
   H^{33}_{\rm{SO}} &= \frac{\sqrt{5} y^4}{8 \sqrt{42} \left(1-e \cos u\right)^4} \Bigg\{\delta  \left(\bm{\hat{\ell}} \cdot \bm{\chi}^{c}_{s}\right) \Bigg[36-132 e^2+126
   e^4-30 e^6-45 \nu +102 e^2 \nu -69 e^4 \nu +12 e^6 \nu + \left(6 e^3  -6 e^5 \right)\, \nonumber \\
   & \times  \cos 3 u+ e^2 \left(1-e^2\right) \cos 2 u \bigg[-20-3
   \nu -6 e^2 \left(1-\frac{\nu }{2}\right)\bigg]- 2 e
   \left(1-e^2\right) \cos u \bigg[-8 \left(1+\frac{9 \nu
   }{4}\right)-17 e^2 \, \nonumber \\
   & \times  \left(1-\frac{18 \nu }{17}\right)\bigg]+ \sqrt{1-e^2} \bigg[\Big(20 i\, e
 -84 i\, e^3  +36 i\, e^5
    -54 i\, e \nu   +98 i\, e^3 \nu 
   -30 i\, e^5 \nu \Big)  \sin u +\Big(24 i\, e^2
   +4 i\, e^4  \, \nonumber \\
   &  +4 i\, e^2 \nu -18 i\, e^4 \nu \Big)  \sin 2 u+\Big(-12 i\,
   e^3  +6 i\, e^3 \nu \Big)  \sin 3 u +\Big(2 i\,
   e^4  - i\, e^4 \nu \Big) \sin 4
   u\bigg]\Bigg] +\left(\bm{\hat{\ell}} \cdot \bm{\chi}^{c}_{a}\right) \Bigg[36-132 e^2  \, \nonumber \\
   &+126 e^4-30 e^6 -171 \nu +606 e^2
   \nu -579 e^4 \nu +144 e^6 \nu +\Big(6 e^3 -6 e^5 -24
   e^3 \nu  +24 e^5 \nu \Big) \cos 3 u+ e^2 \left(1-e^2\right) \, \nonumber \\
   &\times \cos 2 u \bigg[-20+79 \nu -6 e^2 \left(1-\frac{25 \nu
   }{6}\right)\bigg]-2 e \left(1-e^2\right) \cos u \bigg[-8+30 \nu
   -17 e^2 \left(1-\frac{70 \nu }{17}\right)\bigg]+\sqrt{1-e^2} \, \nonumber \\
   &\times \bigg[\Big(20 i\, e 
    -84 i\, e^3 +36 i\, e^5 -90 i\, e \nu  +356 i\, e^3 \nu  
    -154 i\, e^5 \nu\Big)   \sin u+ \Big(24 i\, e^2 
   +4 i\, e^4 -96 i\, e^2 \nu  
   -16 i\, e^4 \nu\Big)  \sin 2 u \, \nonumber \\
   &+\Big(-12 i\, e^3
    +48 i\, e^3 \nu \Big)  \sin 3 u + \Big(2 i\, e^4 -8 i\, e^4 \nu \Big)  \sin 4
   u\bigg]\Bigg]\Bigg\} +\mathcal{O}\left(y^5\right)\,, \\
    H^{32}_{\rm{NS}} &= \frac{ \sqrt{5} \left(1-e^2\right) \left(1-3 \nu\right) y^2}{12 \sqrt{7} \left(1-e \cos u\right)^3}\, \Bigg(4 - 4 e^2 + i\, e
   \sqrt{1-e^2}  \sin u \Bigg) +\mathcal{O}\left(y^4\right)\,, \\
   H^{32}_{\rm{SO}} &= \frac{ \sqrt{5} \left(1-e^2\right) y^3 \nu}{3 \sqrt{7} \left(1-e \cos u\right)^3} \left(\bm{\hat{\ell}} \cdot \bm{\chi}^{c}_{s}\right)  \Bigg(4-4 e^2+ i\, e \sqrt{1-e^2} \sin u\Bigg) +\mathcal{O}\left(y^5\right)\,, \\
   H^{31}_{\rm{NS}} &= \frac{\delta\, y}{12
   \sqrt{14} \left(1-e \cos u\right)^2}  \Bigg\{-1+e^2+ 6 e \left(1-e^2\right) \cos u+\sqrt{1-e^2}\bigg[-6 i\, e
    \sin u+3 i\, e^2  \sin 2 u\bigg]\Bigg\} +\mathcal{O}\left(y^3\right)\,, \\
   H^{31}_{\rm{SO}} &= \frac{y^4}{24 \sqrt{14} \left(1-e \cos u\right)^4}\Bigg\{\delta  \left(\bm{\hat{\ell}} \cdot \bm{\chi}^{c}_{s}\right) \Bigg[4-68 e^2+94 e^4-30 e^6-13 \nu +70 e^2 \nu
   -101 e^4 \nu +44 e^6 \nu +2 e \left(1-e^2\right) \, \nonumber \\
   &\times \bigg[24-14 \nu
   +e^2 \left(1+14 \nu \right)\bigg] \cos u+\left(6 e^3  -6 e^5\right) \cos 3
   u+e^2 \left(1-e^2\right) \cos 2 u \bigg[-20-3 \nu -6 e^2
   \left(1-\frac{\nu }{2}\right)\bigg] \, \nonumber \\
   &+\sqrt{1-e^2}\bigg[\Big(-36 i\, e  -60
   i\, e^3  +12 i\, e^5  +46 i\, e \nu
   -26 i\, e^3 \nu +22 i\, e^5
   \nu  \Big) \sin u+ \Big(72 i\, e^2  +12 i\, e^4
    -36 i\, e^2 \nu -6 i\, e^4
   \nu  \Big) \, \nonumber \\
   & \times \sin 2 u +\Big(-36 i\, e^3  +18 i\,
   e^3 \nu  \Big) \sin 3 u+ \Big(6 i\, e^4 -3 i\,
   e^4 \nu \Big) \sin 4 u\bigg]\Bigg]+\left(\bm{\hat{\ell}} \cdot \bm{\chi}^{c}_{a}\right) \Bigg[4-68 e^2+94
   e^4-30 e^6-11 \nu \, \nonumber \\
   & +254 e^2 \nu  -355 e^4 \nu +112 e^6 \nu + \Big(6 e^3 -6 e^5 -24 e^3 \nu +24 e^5 \nu \Big) \cos 3
   u-2 e \left(1-e^2\right) \cos u \bigg[-24+94 \nu -e^2 \, \nonumber \\
   &\times \left(1-6 \nu
   \right)\bigg] +e^2 \left(1-e^2\right) \cos 2 u \bigg[-20+79 \nu -6 e^2
   \left(1-\frac{25 \nu }{6}\right)\bigg]+\sqrt{1-e^2}\bigg[\Big(-36 i\, e  -60 i\, e^3 +12 i\, e^5 \, \nonumber \\
   &+130 i\, e \nu   +268 i\, e^3 \nu  -62 i\, e^5 \nu \Big) \sin u + \Big(72 i\, e^2   +12 i\, e^4  -288 i\, e^2 \nu  -48 i\, e^4 \nu \Big) \sin 2 u +\Big(-36 i\, e^3 +144 i\, e^3 \nu \Big) \, \nonumber \\
   & \times  \sin 3 u  + \Big(6 i\, e^4
   -24 i\, e^4 \nu \Big)  \sin 4
   u\bigg] \Bigg]\Bigg\} +\mathcal{O}\left(y^5\right)\,, \\
   H^{30}_{\rm{NS}} &= \frac{i}{2 \sqrt{42} }\frac{ e \left(1-e^2\right)^{3/2} (1-3 \nu ) y^2 \sin u}{(1-e \cos u)^3}+\mathcal{O}\left(y^4\right)\,, \\
   H^{30}_{\rm{SO}} &= i\sqrt{\frac{2}{21}}\frac{  e \left(1-e^2\right)^{3/2} \nu \, y^3 \sin u}{(1-e \cos u)^3}\left(\bm{\hat{\ell}} \cdot \bm{\chi}^{c}_{s}\right)+\mathcal{O}\left(y^5\right)\,, \\
   H^{43}_{\rm{NS}} &= -\frac{\left(1-e^2\right)^2 \delta \, \left(1-2 \nu \right)\,y^3}{12 \sqrt{70}
   \left(1-e \cos u\right)^4} \Bigg(27-24 e^2-4 e \cos
   u +e^2 \cos 2 u +10 i\, e \sqrt{1-e^2} \sin u\Bigg) +\mathcal{O}\left(y^5\right)\,, \\
   H^{43}_{\rm{SO}} &= \frac{ \sqrt{5} \left(1-e^2\right)^2 \nu\,y^4}{24 \sqrt{14} \left(1-e \cos u\right)^4} \Bigg[\left(\bm{\hat{\ell}} \cdot \bm{\chi}^{c}_{a}\right)-\delta  \left(\bm{\hat{\ell}} \cdot \bm{\chi}^{c}_{s}\right)\Bigg] \Bigg(27-24 e^2-4 e \cos u +e^2 \cos
   2 u +10 i\, e \sqrt{1-e^2} \sin u\Bigg) \, \nonumber \\
   &+\mathcal{O}\left(y^5\right)\,, \\
   H^{41}_{\rm{NS}} &=-\frac{\left(1-e^2\right)^2 \delta \, \left(1-2 \nu \right)\,y^3}{84
   \sqrt{10} \left(1-e \cos u\right)^4} \Bigg(1+8 e^2-12 e \cos
   u+3 e^2 \cos 2 u +10 i\, e \sqrt{1-e^2} \sin u\Bigg) +\mathcal{O}\left(y^5\right)\,, \\
   H^{41}_{\rm{SO}} &= \frac{\sqrt{5} \left(1-e^2\right)^2 \nu\,y^4 }{168 \sqrt{2} \left(1-e \cos u\right)^4} \Bigg[\left(\bm{\hat{\ell}} \cdot \bm{\chi}^{c}_{a}\right)-\delta  \left(\bm{\hat{\ell}} \cdot \bm{\chi}^{c}_{s}\right)\Bigg] \Bigg(1+8 e^2-12 e \cos u +3 e^2 \cos
   2 u +10 i\, e \sqrt{1-e^2} \sin u\Bigg)\, \nonumber \\
   &+\mathcal{O}\left(y^5\right)\,.
\end{align}
\end{subequations}
\end{widetext}

\bibliographystyle{apsrev4-1}
\bibliography{ref-list}
\end{document}